\documentclass[a4paper,onecolumn,11pt,nopdfoutputerror]{quantumarticle}
\usepackage[utf8]{inputenc}
\usepackage[english]{babel}
\usepackage[T1]{fontenc}
\usepackage{amsmath,amssymb,amsfonts}
\usepackage{mathtools}
\usepackage{graphicx}
\usepackage{hyperref}
\usepackage{algorithm,algpseudocode}

\hypersetup{
    colorlinks=true,
    linkcolor=blue,
    citecolor=blue,
    filecolor=blue,
    urlcolor=blue,
}

\title{Quantum Metropolis--Hastings algorithm}
\author{Jonathan E. Moussa}
\email{godotalgorithm@gmail.com}
\affiliation{Molecular Sciences Software Institute, Virginia Tech, Blacksburg, VA 24060}
\date{}
\begin{document}
\maketitle

\begin{abstract}
I generalize the well-known classical Metropolis--Hastings algorithm into a quantum algorithm that can equilibrate, measure, and mix a quantum thermal state on a quantum computer.
It performs non-symmetric transitions on labels of state preparation and measurement operations and rejects transitions using imprecise energies extracted by Gaussian-filtered quantum phase estimation.
\end{abstract}

\section{Introduction}

Feynman's vision of a quantum computer was of a machine that can simulate quantum many-body systems as efficiently as classical computers can simulate classical many-body systems \cite{quantum_simulation}.
Markov chain Monte Carlo (MCMC) algorithms \cite{mcmc} are the most popular way to simulate classical thermal states on classical computers,
 originating from the work of Metropolis \textit{et al.\@} \cite{metropolis,metropolis_history} and its non-symmetric generalization by Hastings \cite{hastings,hastings_history}.
Quantum MCMC algorithms such as a quantum Metropolis--Hastings algorithm are thus a natural strategy for simulating quantum thermal states on quantum computers.

The first two quantum MCMC algorithms were proposed 25 years ago \cite{qmcmc0}, one based on coupling to a thermal bath and the other based on quantum phase estimation (QPE).
These algorithms prepare quantum thermal states biased by either the finite strength of system-bath coupling or the finite energy resolution of QPE,
 with a characteristic error $\epsilon$ at a cost of Hamiltonian time evolution for duration $O(1/\epsilon)$.
The fundamental appeal of MCMC algorithms is unbiased sampling, which gets expensive to approximate when bias reduction has such poor scaling of cost.
This high cost of bias reduction has persisted in quantum MCMC algorithm development until the recent innovation of filtered QPE with filters chosen to enable bias corrections \cite{qmcmc1,qmcmc2,qmcmc3,qmcmc4}.
Filtered QPE can be approximated to an error $\epsilon$ using Hamiltonian time evolution for duration $O(\log(1/\epsilon))$,
 which effectively eliminates the sampling bias in compatible quantum MCMC algorithms.

In this paper, I generalize the classical Metropolis--Hastings algorithm into a quantum algorithm that emphasizes measurement and uses filtered QPE to correct bias.
I allow for non-symmetric transition probabilities by applying them to labels of state preparation and measurement (SPAM) operations that passively sample thermal expectation values.
State rejection is an iterative process, as in other quantum versions of the Metropolis algorithm \cite{quantum_metropolis}.
However, I use delayed rejection \cite{delayed_rejection} to accept output states with energies that differ from the input state instead of a simple rejection that requires their energies to be equal.
The rejection process has a fundamental stalling problem caused by energy uncertainties and SPAM limitations, which can be repaired by controlled approximations.
This result is a simple baseline for further quantum MCMC algorithm development using a larger set of primitive quantum operations to avoid stalling problems and improve performance.

I have organized the algorithm development in this paper into classical and quantum sections that maximize the number of modifications to the classical Metropolis--Hastings algorithm
 so as to minimize the number of additional modifications needed to develop a quantum algorithm.
This might be a useful approach to adapt other classical algorithms that operate on classical data into effectively equivalent quantum algorithms that operate on quantum data.
The general strategy is to replace primitive classical operations that do not have quantum equivalents with those that do.
In particular, the unnecessary copying of classical data should be avoided since arbitrary quantum data cannot be copied.
This strategy is particularly useful for algorithms with a structure-function relationship where a structure like detailed balance guarantees a function like thermal-state preservation.

The main difference between the quantum Metropolis-Hastings algorithm presented in this paper and other quantum MCMC algorithms that use filtered QPE \cite{qmcmc2,qmcmc3,qmcmc4} is the schedule of measurements.
The quantum Metropolis-Hastings update begins by collapsing a quantum thermal state with a measurement,
 and it lasts a non-deterministic amount of time while a repeat-until-success process attempts to repair the thermal state.
The other algorithms use an MCMC update process that can be unravelled into jump branches and a rejection-like branch.
Jump events formally correspond to measurements of the thermal state, but they occur at non-deterministic times between rejection-like events that do not produce useful measurement information.
Beyond this scheduling difference, all of these algorithms will have different cost prefactors
 that depend on details such as mixing time relative to a base cost of $O(\log(1/\epsilon))$ Hamiltonian evolution time per measurement.

\section{Classical algorithm\label{classical_section}}

I begin by reviewing the classical Metropolis--Hastings algorithm before considering other algorithms in subsequent sections on the path to a practical quantum algorithm.

The Metropolis--Hastings algorithm generates a sequence of samples from a finite set of states $\mathcal{S}$ by applying Algorithm \ref{MH} to generate each sample from the previous sample in the sequence.
This random process is a time-homogeneous Markov chain defined by the conditional probability function
\begin{equation} \label{MH_probability}
 P_M(b|a) = A(b,a) P(b|a) + \delta_{a,b} \sum_{c\in \mathcal{S}}[1 - A(c,a)] P(c|a)
\end{equation}
 of generating state $b$ from the previous state $a$, where $\delta_{a,b}$ is the Kronecker delta function, $P$ is a driving conditional probability function that must be sampled and evaluated,
 and $A$ is the Metropolis--Hastings probability for accepting a state transition,
\begin{equation} \label{accept}
 A(b,a) = \min \{1, \exp[\beta E(a) - \beta E(b)] P(a|b)/P(b|a) \}
\end{equation}
  for a reciprocal temperature $\beta$ and energy function $E$ that define a thermal distribution,
\begin{equation} \label{thermal}
 p(a) = \frac{\exp[-\beta E(a)]}{Z} \ \ \mathrm{for} \ \ Z = \sum_{a \in \mathcal{S}} \exp[-\beta E(a)].
\end{equation} 
Eq.\@ (\ref{MH_probability}) can be derived from Algorithm \ref{MH} by summing the conditional probability of each statistical branch corresponding to the different sequences of random outcomes
 and then evaluating each conditional probability as the product of probabilities for each outcome in that branch.
The first term in Eq.\@ (\ref{MH_probability}) corresponds to the \textbf{if} condition being true and accepting the new state $b$, and the remaining terms for $b = a$ correspond to rejecting the new state $c$ and retaining the old state $a$.
Note that $A(b,a)$ does not need to be well defined for $a,b \in \mathcal{S}$ such that $P(b|a) = 0$ because those events cannot occur in Algorithm \ref{MH}.

\begin{algorithm}[t]
\caption{Classical Metropolis--Hastings update\label{MH}}
\begin{algorithmic}[1]
\Require{state $a \in \mathcal{S}$, energy $E: \mathcal{S} \rightarrow \mathbb{R}$, reciprocal temperature $\beta \in \mathbb{R}_{\ge 0}$, probability distribution $P$ over $\mathcal{S}$ conditioned on $\mathcal{S}$}
\Ensure{$a$ is replaced by a sample from $P_M(\cdot|a)$ defined by Eqs.\@ (\ref{MH_probability}) and (\ref{accept})}
\State $b \gets \textrm{sample from} \ P(\cdot | a)$
\State $u \gets \textrm{sample from the uniform distribution over} \ [0,1]$
\If{$\displaystyle u \le \exp[\beta E(a) - \beta E(b)] P(a|b)/P(b|a)$}
\State $a \gets b$
\EndIf
\end{algorithmic}
\end{algorithm}

The essential property of $P_M$ is that it satisfies the detailed balance condition,
\begin{equation} \label{balance}
 P_M(b|a) \exp[-\beta E(a)] = P_M(a|b) \exp[-\beta E(b)],
\end{equation}
 which is independently satisfied by each term in the sum on the right side of Eq.\@ (\ref{MH_probability}) that correspond to the acceptance and rejection branches.
Detailed balance guarantees that $p$ is a stationary distribution of $P_M$,
\begin{equation} \label{stationary}
 \sum_{a\in\mathcal{S}} P_M(b|a) p(a) = p(b).
\end{equation}
Thus, if the first sample in the Markov chain is drawn from $p$, then all subsequent samples also have $p$ as their marginal distribution.
Eq.\@ (\ref{stationary}) follows from Eq.\@ (\ref{balance}) by summing over $a$, dividing by $Z$ on both sides of the equation,
 and removing $P_M(a|b)$ from the right side because the total probability over all outcomes is always one.

Typically, a Markov chain is the only way to sample from $p$ in Eq.\@ (\ref{thermal}) efficiently, and this chain is initialized by sampling from another distribution $p_0$ for which independent samples can be generated efficiently.
The marginal distribution after $n$ steps is then
\begin{equation}
 p_n(b) = \sum_{a \in \mathcal{S}} P_M(b|a) p_{n-1}(a).
\end{equation}
After enough steps of the Markov chain, these marginal distributions will equilibrate and approximate $p$.
The total variation distance between $p_n$ and $p$ decays exponentially as
\begin{equation} \label{eq_error}
 \frac{1}{2}\sum_{a \in \mathcal{S}} | p_n(a) - p(a) | \le \Omega_M^n \frac{1}{2} \sum_{a \in \mathcal{S}} | p_0(a) - p(a) |
\end{equation}
 for $n \ge 1$ and a maximum retention rate $\Omega_M$ of $P_M$.
Generically, the maximum retention rate $\Omega_*$ for a conditional probability $P_*$ is defined as
\begin{equation} \label{mix_time}
 \Omega_* = \max_{x} \frac{\sum_{b \in \mathcal{S}} \left| \sum_{a \in \mathcal{S}} P_*(b|a) x(a) \right| }{\sum_{a \in \mathcal{S}} |x(a)|} \ \ \ \mathrm{subject \ to} \ \ \ \sum_{a \in \mathcal{S}}x(a) = 0.
\end{equation}
The upper bound in Eq.\@ (\ref{eq_error}) is derived by rewriting the target distance, $\epsilon_n$, as a product of ratios, $(\epsilon_n/\epsilon_{n-1}) \cdots (\epsilon_1/\epsilon_0) \epsilon_0$,
 and bounding $\epsilon_m/\epsilon_{m-1} \le \Omega_M$ for all $m \ge 1$ since $\Omega_M$ in Eq.\@ (\ref{mix_time}) corresponds to the ratios maximized over differences between pairs of probability distributions.
Eq.\@ (\ref{mix_time}) is a variational characterization of the second largest eigenvalue of $P_*$ by maximizing over the subspace orthogonal to the left eigenvector of the guaranteed eigenvalue one,
 with all of the vector elements of this eigenvector being equal to one.

Once the Markov chain is sufficiently equilibrated, each subsequent step generates an unbiased sample from $p$ that is correlated with the previous sample.
With enough steps between samples, their joint distribution will mix and approximate independent samples.
Specifically, the pair distribution of samples separated by $n$ steps of the Markov chain is
\begin{equation} \label{mix_state}
 p_0^{\mathrm{pair}}(a,b) = p(a) \delta_{a,b}, \ \ \ p_n^{\mathrm{pair}}(a,b) = \sum_{c \in \mathcal{S}} P_M(a|c) p_{n-1}^{\mathrm{pair}}(c,b),
\end{equation}
 and its total variation distance from the pair distribution of independent $p$ samples is
\begin{equation} \label{mix_error}
 \frac{1}{2}\sum_{a,b \in \mathcal{S}} | p_n^{\mathrm{pair}}(a,b) - p(a) p(b) | \le \Omega_M^n \left(1 - \sum_{a \in \mathcal{S}} p(a)^2 \right)
\end{equation}
 for $n \ge 1$.
This bound has a similar derivation as Eq.\@ (\ref{eq_error}), with an extra simplification of the distance on the right side
 and an extra complication that the maximization over pair probabilities, factored as $q(b|a) p(a)$, has to be further relaxed with an extra step as
\begin{align} \label{classical_pair_proof}
 \frac{\sum_{a,b \in \mathcal{S}}| p_n^{\mathrm{pair}}(a,b) - p(a) p(b) |}{\sum_{a,b \in \mathcal{S}}| p_{n-1}^{\mathrm{pair}}(a,b) - p(a) p(b) |}
 &= \frac{\sum_{a,b \in \mathcal{S}}| \sum_{c \in \mathcal{S}} P_M(a|c) [p_{n-1}^{\mathrm{pair}}(c|b) - p(c)]  | p(b)}{\sum_{a,b \in \mathcal{S}}| p_{n-1}^{\mathrm{pair}}(a|b) - p(a) | p(b)} \notag \\
 &\le \sum_{b \in \mathcal{S}} p(b) \frac{\sum_{a \in \mathcal{S}}| \sum_{c \in \mathcal{S}} P_M(a|c) [p_{n-1}^{\mathrm{pair}}(c|b) - p(c)]  |}{\sum_{a \in \mathcal{S}}| p_{n-1}^{\mathrm{pair}}(a|b) - p(a) |}\notag \\
 & = \Omega_M ,
\end{align}
 using Chebyshev's sum inequality to bound the ratio of averages by the average of ratios and that Eq.\@ (\ref{mix_state}) preserves $p$ as the marginal distribution of $p_n^{\mathrm{pair}}(a,b)$ over $b$.

Besides $\Omega_M$, another important performance property is the average rejection rate $\Lambda$ when the \textbf{if} condition is true in Algorithm \ref{MH},
\begin{equation} \label{reject_rate}
 \Lambda = 1 - \sum_{a,b \in \mathcal{S}} A(b,a) P(b|a) p(a).
\end{equation}
It is equal to the total variation distance between the pair distribution of sampling from $p$ then sampling from $P$ conditioned on the outcome and its time-reversed distribution,
\begin{equation} \label{reject_equiv}
 \Lambda = \frac{1}{2} \sum_{a,b \in \mathcal{S}} | P(b|a) p(a) - P(a|b) p(b) |,
\end{equation}
 with the absolute value here related to the minimum in Eq.\@ (\ref{reject_rate}) via Eq.\@ (\ref{accept}) through the identity $\min \{ x, y \} = (x+y)/2 - |x-y|/2$.
Thus $\Lambda$ is a direct measure of the reversibility of $P$ with respect to $p$.
Also, the asymptotic distribution of trial updates in Algorithm \ref{MH},
\begin{equation}
 p_{\mathrm{try}}(b) = \sum_{a \in \mathcal{S}} P(b|a) p(a),
\end{equation}
 is bounded in total variation distance from $p$ by $\Lambda$,
\begin{equation} \label{try_dist}
  \frac{1}{2} \sum_{a \in \mathcal{S}} | p_{\mathrm{try}}(a) - p(a) | \le \Lambda,
\end{equation}
 which follows from applying the triangle inequality to the sum over $a$ in Eq.\@ (\ref{reject_equiv}), to move it inside the absolute value and eliminate $P(a|b)$.

Since Algorithm \ref{MH} does not change the state after a rejection, a high $\Lambda$ should imply a high $\Omega_M$.
To show this, I reorganize the acceptance and rejection branches in Eq.\@ (\ref{MH_probability}),
\begin{equation}
 P_M(b|a) = P_A(b|a) [1 - \lambda(a)] + \delta_{a,b} \lambda(a),
\end{equation}
 as a rejection probability $\lambda$ and a conditional probability $P_A$ after acceptance,
\begin{equation}
  \lambda(a) = \sum_{b \in \mathcal{S}}[1 - A(b,a)] P(b|a) \ \ \ \mathrm{and} \ \ \ P_A(b|a) = \frac{A(b,a) P(b|a)}{1 - \lambda(a)},
\end{equation}
 where the expectation value of $\lambda$ with respect to $p$ is $\Lambda$.
If Algorithm \ref{MH} is repeated until a new state is accepted up to $n \ge 1$ repetitions, then the total conditional probability is
\begin{equation} \label{acceptance_split}
 P_M^{\le n}(b|a) = P_A(b|a) [1 - \lambda(a)] \sum_{m=0}^{n-1} \lambda(a)^m  + \delta_{a,b} \lambda(a)^n.
\end{equation}
If $\lambda(a) < 1$ for all $a \in \mathcal{S}$, then $\mathrm{lim}_{n \rightarrow \infty}P_M^{\le n} = P_A$.
With a rejection rate of $\Lambda$, Algorithm \ref{MH} needs to be run $1/(1-\Lambda)$ times on average to perform $P_A$.
The retention rate $\Omega_A$ of $P_A$ defined by Eq.\@ (\ref{mix_time}) is then approximately related to $\Omega_M$ by
\begin{equation} \label{approx_relation}
 \Omega_M \approx \Omega_A^{1- \Lambda},
\end{equation}
 which would be exact if the distribution of runs to implement $P_A$ was fully concentrated on $1/(1-\Lambda)$.
As expected, a high $\Lambda$ results in a high $\Omega_M$, but a low $\Lambda$ does not guarantee a low $\Omega_M$ unless $\Omega_A$ is low, which depends on the choice of $P$.

I further illustrate the relationship between $\Lambda$ and $\Omega_M$ with a simple example.
For any thermal distribution $p$ over any set of states $\mathcal{S}$, I consider an extension to $\mathcal{S}' = \mathcal{S} \cup \{ \infty \}$ and $p(\infty) = 0$,
 and a two-parameter conditional probability function,
\begin{align}
P'(b|a;\Lambda', \Omega') &= (1-\Omega') p'(b ; \Lambda') + \Omega' \delta_{a,b}, \notag \\
p'(a ; \Lambda') &= (1 - \Lambda') p(a) + \Lambda' \delta_{a,\infty}, \label{MH_example}
\end{align}
 with a stationary state $p'$ and a retention rate $\Omega'$ defined by Eq.\@ (\ref{mix_time}).
When $P'$ is used in Algorithm \ref{MH}, the conditional probability is
\begin{equation} \label{M_example}
 P'_M(b|a;\Lambda', \Omega') = [1-\Omega'_M(\Lambda' , \Omega')] p(b) + \Omega'_M(\Lambda', \Omega') \delta_{a,b},
\end{equation}
 with a rejection rate of $\Lambda'$ that saturates the bound in Eq.\@ (\ref{try_dist}) and a retention rate of
\begin{equation} \label{simple_relation}
 \Omega'_M(\Lambda', \Omega') = 1 - (1- \Omega') (1-\Lambda').
\end{equation}
The conditional probability after acceptance as defined by Eq.\@ (\ref{acceptance_split}) has the same form as in Eq.\@ (\ref{M_example}) with $\Omega'$ instead of $\Omega'_M$.
Eqs. (\ref{approx_relation}) and (\ref{simple_relation}) are consistent for $\Omega_A = \Omega' \approx 1$.

\subsection{Practical considerations\label{classical_practical}}

In practice, Algorithm \ref{MH} is two distinct algorithms with different design spaces.
A general $P$ requires the generation of one sample and two evaluations of its conditional probability function to run Algorithm \ref{MH}.
A symmetric $P$ that satisfies $P(b|a) = P(a|b)$ for all $a,b \in \mathcal{S}$ only requires sample generation from $P$ to run Algorithm \ref{MH},
 because $P$ cancels from the acceptance probability in Eq.\@ (\ref{accept}).
Thus, the symmetry constraint on $P$ is paired with the freedom from evaluating its conditional probability function.
This allows $P$ to be defined implicitly as a sampling process with a proof of symmetry, like in the original Metropolis algorithm \cite{metropolis}.
It is not obvious that a symmetric, evaluation-free $P$ is more cost effective than a general $P$ at generating samples from $p$ for a given $\mathcal{S}$, $E$, and $\beta$.

The difference between the two design spaces is illustrated by optimal but impractical choices of $P$.
For a non-symmetric $P$, the uniquely optimal choice is $P(b|a) = p(b)$, which results in $P = P_M = P_A$ with rejection and retention rates of zero.
Sampling from such a $P$ corresponds to independently sampling from $p$, but an MCMC algorithm is not needed if $p$ can be sampled efficiently and accurately.
The same optimal $P_A$ with a retention rate of zero can be generated by a symmetric $P$ that is easy to sample, $P(b|a) = 1/|\mathcal{S}|$,
 but with a high rejection rate bounded from below in Eq.\@ (\ref{try_dist}) by the total variation distance between $p$ and the uniform distribution over $\mathcal{S}$.
It is still possible to reduce this rejection rate by modifying $\mathcal{S}$ and $E$.
Specifically, $\mathcal{S}$ can be embedded in a larger set of states $\mathcal{S}_n$ by introducing an internal set of states $\mathcal{I}_n(a)$ for each $a \in \mathcal{S}$ such that
\begin{equation}
 \mathcal{S}_n = \{ (a,i) : a \in \mathcal{S}, i \in \mathcal{I}_n(a) \} , \ \ \ |\mathcal{I}_n(a)| = \left\lfloor n \exp\left(\max_{b \in \mathcal{S}} \beta E(b) -\beta E(a)\right) \right\rfloor ,
\end{equation}
 and then defining the state energy on the expanded domain $\mathcal{S}_n$ as
\begin{equation}
 E_n(a,i) = E(a) + \beta^{-1} \log |\mathcal{I}_n(a)|.
\end{equation}
This embedding defines new stationary distributions that all preserve $p$ as their marginal distribution defined by the map $(a,i) \mapsto a$ from $\mathcal{S}_n$ to $\mathcal{S}$.
The rejection rate approaches zero as $n$ increases and the embedded thermal distribution becomes increasingly uniform.
This embedding is not efficient to construct in general, but other embedding methods are used in practice to improve the efficiency of MCMC sampling.
A popular example is the Swendsen--Wang algorithm \cite{swendsen_wang} that embeds spin models into a larger configuration space that allows coupled spins to be grouped into clusters of aligned spins.

Another common practice is to generate a time-inhomogeneous Markov chains defined by repeating Algorithm \ref{MH} for multiple choices of $P$ but a common choice of $\mathcal{S}$, $E$, and $\beta$.
This expands the design space relative to a single choice of $P$ while still asymptotically generating samples from $p$, but its behavior is more difficult to analyze.
If a sequence of $n$ choices of $P$ is repeated, then subsequences of samples from the Markov chain
 separated by $n-1$ samples are still time-homogeneous Markov chains.
The conditional probability of these subsequences can then be analyzed using Eqs.\@ (\ref{eq_error}) and (\ref{mix_error})
 and described by an overall maximum retention rate $\Omega$ and average rejection rate $\Lambda$.

For the analysis in this section to be useful in practice, $\Lambda$ and $\Omega_M$ should be efficient to estimate.
Since $\Lambda$ is an expectation value of a two-outcome distribution of truth values for the \textbf{if} condition in Algorithm \ref{MH}, it can be estimated efficiently while sampling from the Markov chain.
The direct evaluation of $\Omega_M$ from Eq.\@ (\ref{mix_time}) is not efficient for practical applications of MCMC algorithms with large values of $|\mathcal{S}|$.
In practice, it is more efficient to underestimate $\Omega_M$ by analyzing the decay of observed correlations in the Markov chain.
For example, $\mathcal{S}$ can be reduced to a smaller set of observations $\mathcal{\overline{S}}$ by a function $f: \mathcal{S} \rightarrow \mathcal{\overline{S}}$.
The asymptotic coarse-grained Markov process and its stationary distribution over $\mathcal{\overline{S}}$ are
\begin{equation}\label{coarse}
 \overline{P}_M(\overline{b}|\overline{a}) = \sum_{b \in f^{-1}(\overline{b})} \sum_{a \in f^{-1}(\overline{a})} P_M(b|a) p(a)/\overline{p}(f(a)) \ \ \ \mathrm{and} \ \ \ \overline{p}(\overline{a}) = \sum_{a \in f^{-1}(\overline{a})} p(a),
\end{equation}
 where $f^{-1}(\overline{a}) \subseteq \mathcal{S}$ refers to the preimage of $\overline{a} \in \mathcal{\overline{S}}$ with respect to $f$.
For sufficiently small values of $|\mathcal{\overline{S}}|$, $\overline{P}_M$ can be estimated directly with samples from the Markov chain,
 and its retention rate $\overline{\Omega}_M$ can then be calculated from Eq.\@ (\ref{mix_time}).
The definition of $\overline{\Omega}_M$ is equivalent to variationally restricting the maximization in the definition of $\Omega_M$ from $x$ to $\overline{x}$ as
\begin{equation}
 x(a) = p(a) \overline{x}(f(a)) / \overline{p}(f(a)),
\end{equation}
 resulting in an underestimate of the true retention rate, $\overline{\Omega}_M \le \Omega_M$.
Overestimating the amount of equilibration and mixing performed by $P_M$ should be expected in general from any incomplete analysis that can overlook retained correlations.
A similar coarse-grained analysis can be used for $P_A$ by postselecting state transitions from the acceptance branch of $P_M$,
 although the number of samples will then be reduced by a factor of $1-\Lambda$.

\section{Imprecise algorithm\label{imprecise_section}}

Now I consider how the Metropolis--Hastings algorithm can be adapted to a limited ability to control and measure physical states.
This section considers a classical algorithm with artificial limitations that mirror the natural limitations of a quantum algorithm, and the results are then translated into a quantum algorithm in the next section.

In a quantum version of Algorithm \ref{MH}, it will not be possible to read or write the states $a$ and $b$ directly without collapsing the quantum state and introduce biasing errors.
This problem can be avoided by instead considering a finite set of observations $\mathcal{O}$ that can be accessed by the computer.
The computer can then indirectly measure an unknown state through an observation process that returns an element of $\mathcal{O}$ and also modifies the state.
This process is defined by a probability distribution $P_O$ over $\mathcal{O} \times \mathcal{S}$ that is conditioned on $\mathcal{S}$.
Similarly, the computer manipulates the state through a control process defined by a probability distribution $P_C$ over $\mathcal{S}$ that is conditioned on $\mathcal{O} \times \mathcal{S}$
 for a known observation and an unknown state.
The only requirement on $P_O$ and $P_C$ is a symmetry condition,
 \begin{equation} \label{classical_spam}
 P_S(i, b | o, a) = \sum_{c \in \mathcal{S}} P_C(b|o,c) P_O(i,c|a) = P_S(o, a | i, b),
\end{equation}
 for all $a,b \in \mathcal{S}$ and $i,j \in \mathcal{O}$, which avoids the need for detailed knowledge of $P_O$ and $P_C$.
Direct state access then corresponds to $\mathcal{O} = \mathcal{S}$, $P_C(b|i,a) = \delta_{b,i}$, and $P_O(i,b|a) = \delta_{i,a} \delta_{b,a}$.
Energy measurements are special and do not collapse quantum thermal states, but they can only be estimated with finite precision in practical quantum algorithms.
In a classical algorithm, this imprecision corresponds to measuring the energy of an unknown state $a$ by sampling from the normal distribution with mean $E(a)$ and variance $\sigma^2$.

\begin{algorithm}[t]
\caption{Imprecise Metropolis--Hastings update \label{IMH}}
\begin{algorithmic}[1]
\Require{ $n_{\max} \in \mathbb{N}$, state $a \in \mathcal{S}$, reciprocal temperature $\beta \in \mathbb{R}_{\ge 0}$,  energy uncertainty $\sigma \in \mathbb{R}_{\ge 0}$,
 energy $E: \mathcal{S} \rightarrow \mathbb{R}$,
 probability distributions $P_C$ over $\mathcal{S}$ conditioned on $\mathcal{O} \times \mathcal{S}$,
 $P_O$ over $\mathcal{O} \times \mathcal{S}$ conditioned on $\mathcal{S}$ satisfying Eq.\@ (\ref{classical_spam})},
 $P$ over $\mathcal{O}$ conditioned on $\mathcal{O}$
\Ensure{$a$ is replaced by a sample from $\widetilde{P}_I(\gamma, \cdot|a)$ defined by Eqs.\@ (\ref{classical_sym})--(\ref{IMH_probability2}) and (\ref{simple_accept})}
\State $\omega_0 \gets$ sample from the normal distribution with mean $E(a)$ and variance $\sigma^2$
\State $(i, a) \gets \textrm{sample from} \ P_O(\cdot, \cdot | a)$
\State $o \gets \textrm{sample from} \ P(\cdot | i)$
\State $a \gets \textrm{sample from} \ P_C(\cdot | o, a)$
\State $x_{\min} \gets \exp(-\beta \omega_0 + \beta^2 \sigma^2) P(o|i)$
\State $\omega \gets$ sample from the normal distribution with mean $E(a)$ and variance $\sigma^2$
\State $x \gets x_{\min} - \exp(- \beta \omega) P(i|o)$
\State $u \gets \textrm{sample from the uniform distribution over} \ [0,1]$
\State $ \gamma \gets (i, \omega, o, \omega_0)$
\State $(o, n) \gets (i, 1)$
\While{$x > x_{\min} u \ \mathrm{and} \ n < n_{\max}$}
\State $(i, a) \gets \textrm{sample from} \ P_O(\cdot, \cdot | a)$
\State $a \gets \textrm{sample from} \ P_C(\cdot | o, a)$
\State $x_{\min} \gets \min\{ x_{\min} , x \}$
\State $x \gets x + \exp(- \beta \omega) P(o|i)$
\State $\omega \gets$ sample from the normal distribution with mean $E(a)$ and variance $\sigma^2$
\State $x \gets x - \exp(- \beta \omega) P(i|o) $
\State $u \gets \textrm{sample from the uniform distribution over} \ [0,1]$
\State $ \gamma \gets (i, \omega, \gamma)$
\State $(o, n) \gets (i, n+1)$
\EndWhile
\end{algorithmic}
\end{algorithm}

Like Algorithm \ref{MH} in the previous section, Algorithm \ref{IMH} is again a random process that generates a time-homogeneous Markov chain defined by a conditional probability function,
 but it now has a more complicated branch structure.
First, I consider the ideal algorithm in the $n_{\max} \rightarrow \infty$ limit and use measure theory to describe the branches compactly as
\begin{equation} \label{IMH_probability}
 P_I(b|a) = \int_{\mathcal{M}} d\mu(\gamma) P_I(\gamma, b| a), \ \ \ P_I(\gamma_n, b | a) = \sum_{\mathbf{s} \in \mathcal{S}^{n-1}} P_I(\gamma_n, b, \mathbf{s} | a),
\end{equation}
 for $\gamma_{n} \in \mathcal{M}_{n}$, which is marginalized over an unknown outcome in $\mathcal{S}^{n-1}$ for $n \ge 2$
 and a known outcome in $\mathcal{M} = \bigcup_{n = 1}^{\infty} \mathcal{M}_n$ for $\mathcal{M}_n = (\mathcal{O} \times \mathbb{R})^{n+1}$.
Here, Lebesgue integrals and measure theory are only being used as a compact notation for standard sums and integrals.
For this purpose, $\mu$ is a product measure that combines the counting measure on $\mathcal{O}$ and the Lebesgue measure on $\mathbb{R}$, which respectively correspond to sums and integrals.
Every imprecise energy measurement and known $P_O$ outcome is combined to form the outcome $\gamma \in \mathcal{M}$ of $P_I$.
Algorithm \ref{IMH} with $\sigma = 0$, $n_{\max} \ge 2$, and $P_C$ and $P_O$ that implement direct state access is equivalent to Algorithm \ref{MH}.

The conditional probability function over all outcomes in Eq.\@ (\ref{IMH_probability}) is a product of event probabilities that can be split into symmetric and decision components,
\begin{equation} \label{classical_split}
 P_I(\gamma_n, a_n, \cdots , a_1 | a_0) = P_I^{\mathrm{sym}}(\gamma_n, a_n, \cdots , a_1 | a_0) P_I^{\mathrm{dec}}(\gamma_n),
\end{equation}
 where $a_n$ is the output of the $n$th instance of $P_C$ and the input to the $(n+1)$th instance of $P_O$.
The symmetric component contains the normal distributions of the imprecise energy measurements and $P_C$ and $P_O$ marginalized over their intermediate state,
\begin{equation} \label{classical_sym}
 P_I^{\mathrm{sym}}(\gamma_n, a_n, \cdots , a_1 | a_0) = \left( \prod_{m=0}^n  \frac{e^{-[\omega_m-E(a_m)]^2/(2\sigma^2)}}{\sqrt{2 \pi} \sigma} \right) \prod_{m=0}^{n-1} P_S(o_{m+1}, a_{m+1} | o_m, a_m),
\end{equation}
 for $\gamma_n = (o_n, \omega_n, \cdots, o_0, \omega_0) \in \mathcal{M}_n$.
I avoid defining explicit elements for $\gamma \in \mathcal{M}$ by using indexing functions $\underline{\omega}_n : \mathcal{M} \rightarrow \mathbb{R}$
 and $\underline{o}_n : \mathcal{M} \rightarrow \mathcal{O}$ such that $\underline{\omega}_n(\cdots, o_1, \omega_1, o_0, \omega_0) = \omega_n$ and $\underline{o}_n(\cdots, o_1, \omega_1, o_0, \omega_0) = o_n$.
A recursive form for the decision component is then
\begin{align} \label{R_recursion}
 P_I^{\mathrm{dec}}(\gamma) &= P(\underline{o}_0(\gamma) | \underline{o}_1(\gamma)) A(\gamma) R(\gamma), \notag \\ 
 R(o, \omega, o', \omega') &= 1, \ \ \ R(o, \omega, \gamma) = [ 1 - A(\gamma) ] R(\gamma),
\end{align}
 where $A(\gamma)$ is the probability of accepting the outcome $\gamma$ and $R(\gamma)$ is the probability of rejecting all branches prior to $\gamma$.
For finite values of $n_{\max}$, Algorithm \ref{IMH} is described by
\begin{align} \label{IMH_probability2}
\widetilde{P}_I(\gamma_n, b | a) &= \sum_{\mathbf{s} \in \mathcal{S}^{n-1}} P_I^{\mathrm{sym}}(\gamma_n, b, \mathbf{s} | a) \widetilde{P}_I^{\mathrm{dec}}(\gamma_n), \notag \\
\widetilde{P}_I^{\mathrm{dec}}(\gamma_n) &= \left\{ \begin{array}{ll} P_I^{\mathrm{dec}}(\gamma_n), & n < n_{\max}  \\ 
 P(\underline{o}_0(\gamma_n) | \underline{o}_1(\gamma_n)) R(\gamma_n), & n = n_{\max}  \\ 
 0, & n > n_{\max} \end{array} \right. ,
\end{align}
 which exactly matches the behavior of $P_I$ for $n < n_{\max}$ and approximates the behavior of $P_I$ for $n = n_{\max}$ with an error proportional to $1- A(\gamma)$ for a terminal outcome $\gamma$.

Following the structure of delayed rejection \cite{delayed_rejection}, I enforce detailed balance on $P_I$ as a whole by requiring detailed balance between pairs of branches in Eq.\@ (\ref{IMH_probability}).
The selection of these pairs must carefully consider the marginalization of variables to achieve a satisfiable balance condition.
I begin with the overall detailed balance condition from Eq.\@ (\ref{balance}),
\begin{equation} \label{branch_balance}
\int_{\mathcal{M}} d\mu(\gamma) P_I(\gamma, b| a) e^{-\beta E(a)} = \int_{\mathcal{M}} d\mu(\gamma) P_I(\gamma, a| b) e^{-\beta E(b)},
\end{equation}
 which is difficult to satisfy if the energies $E(a)$ and $E(b)$ are not known precisely for the Metropolis--Hastings acceptance probability function in Eq.\@ (\ref{accept}).
The precise energies can be replaced with imprecise energies by using a Gaussian integral identity
\begin{equation} \label{gaussian_identity} 
 \int_{\mathbb{R}} d\omega e^{-(\omega - E)^2/(2 \sigma^2)} e^{-\beta E} f(\omega) = e^{-\beta^2 \sigma^2/2} \int_{\mathbb{R}} d\omega e^{-(\omega - E)^2/(2 \sigma^2)} e^{-\beta \omega} f(\omega + \beta \sigma^2)
\end{equation}
 for any function $f$, since these Gaussians are a term in the branch probabilities in Eq.\@ (\ref{classical_sym}).
Using this identity, I rewrite Eq.\@ (\ref{branch_balance}) with imprecise Boltzmann weights as
\begin{equation} \label{branch_balance2}
\int_{\mathcal{M}} d\mu(\gamma) P_I(\gamma + \beta \sigma^2, b| a) e^{-\beta \underline{\omega}_0(\gamma)} = \int_{\mathcal{M}} d\mu(\gamma) P_I(\gamma + \beta \sigma^2, a| b) e^{-\beta \underline{\omega}_0(\gamma)},
\end{equation}
 where additions between a scalar and a tuple are applied to the right-most element of the tuple.
This condition can be satisfied by balancing time-reversed pairs of branches,
\begin{equation} \label{branch_balance3}
 P_I(\gamma + \beta \sigma^2, a_n, \cdots , a_1 | a_0) e^{-\beta \underline{\omega}_0(\gamma)} = P_I(\overline{\gamma} + \beta \sigma^2, a_0, \cdots , a_{n-1} | a_n) e^{-\beta \underline{\omega}_0(\overline{\gamma})},
\end{equation}
 where $\overline{\gamma}$ denotes the time-reversed version of $\gamma \in \mathcal{M}$ as
 defined by the measure-preserving automorphism, $(o_n, \omega_n, \cdots, o_0, \omega_0) \mapsto (o_0, \omega_0, \cdots, o_n, \omega_n)$.
This specific choice of branch pairing allows equal $P_I^{\mathrm{sym}}$ terms to be removed from both sides of Eq.\@ (\ref{branch_balance3}), leaving
\begin{align} \label{branch_balance4}
P(\underline{o}_0(\gamma)| \underline{o}_1(\gamma)) (A \cdot R)(\gamma + \beta \sigma^2) e^{-\beta \underline{\omega}_0(\gamma)} = P(\underline{o}_0(\overline{\gamma})| \underline{o}_1(\overline{\gamma})) (A \cdot R)(\overline{\gamma} + \beta \sigma^2) e^{-\beta \underline{\omega}_0(\overline{\gamma})},
\end{align}
 which is a sufficient condition for Eq.\@ (\ref{branch_balance}) that does not contain elements of $\mathcal{S}$.

Just as in Eqs.\@ (\ref{accept}) and (\ref{balance}), the acceptance probabilities are chosen to satisfy Eq.\@ (\ref{branch_balance4}).
The first acceptance probability is a bias-corrected version of Eq.\@ (\ref{accept}),
\begin{equation}
 A(o_1, \omega_1, o_0, \omega_0) = \min \{1, \exp(\beta \omega_0 - \beta \omega_1 - \beta^2 \sigma^2) P(o_1|o_0)/P(o_0|o_1) \}.
\end{equation}
This $\beta^2 \sigma^2$ bias correction for absolute energy measurements with variance $\sigma^2$
 is similar to a prior $\beta^2 \sigma^2/2$ bias correction for relative energy measurements with variance $\sigma^2$ \cite{uncertain_metropolis}.
In general, Eq.\@ (\ref{branch_balance4}) can be satisfied by a recursive formula for acceptance probabilities,
\begin{equation} \label{generic_accept}
 A(\gamma + \beta \sigma^2) = \min \left\{ 1, \frac{e^{-\beta \underline{\omega}_0(\overline{\gamma})} P(\underline{o}_0(\overline{\gamma})| \underline{o}_1(\overline{\gamma})) R(\overline{\gamma} + \beta \sigma^2)}{e^{-\beta \underline{\omega}_0(\gamma) }P(\underline{o}_0(\gamma)| \underline{o}_1(\gamma)) R(\gamma + \beta \sigma^2)} \right\},
\end{equation}
 which is analogous to the general form of delayed rejection \cite{delayed_rejection} for $\sigma = 0$.
As derived in Appendix \ref{acceptance_probability}, this recursive formula can be solved and replaced by an explicit formula,
\begin{align} \label{simple_accept}
 &A(o_n, \omega_n, \cdots , o_1, \omega_1, o_0, \omega_0 + \beta \sigma^2) = \min \left\{ 1, \frac{\max \{ 0,  \min_{r \in \{ 0, \cdots, n-1 \}} x_r - x_n \} } {\max \{ 0,  \min_{r \in \{ 0, \cdots, n-1 \}} x_r \} } \right\}, \notag \\ &x_0 =  e^{-\beta \omega_0} P(o_0|o_1), \ \ \ x_n =  \sum_{r=0}^{n-1} [ e^{-\beta \omega_r} P(o_r|o_{r+1}) - e^{-\beta \omega_{r+1}} P(o_{r+1}|o_r) ] .
\end{align}
Algorithm \ref{IMH} stores the denominator of $A$ in $x_{\min}$ and accumulates new values of $x_n$ in $x$.
Similar to Eq.\@ (\ref{accept}), $A$ may be undefined if its denominator is zero, but this is irrelevant to detailed balance because it corresponds to a sequence of outcomes with zero probability.
The symmetric form of delayed rejection \cite{delayed_rejection} corresponds to the choice of $P(o|i) = 1/|\mathcal{O}|$, which allows $x_n$ to be simplified from partial sums to pairs of Boltzmann weights,
\begin{equation}
  x_0 =  e^{-\beta \omega_0}, \ \ \ x_n =  e^{-\beta \omega_0} - e^{-\beta \omega_n}.
\end{equation}
Unlike in Algorithm \ref{MH}, a more general symmetric choice of $P$ cannot be removed from the acceptance probability.
Instead, $P_S$ serves as a symmetric component of the state update process that does not need to be evaluated.
However, $P_S$ is sampled once per loop of the delayed rejection process whereas $P$ is sampled only once per run of Algorithm \ref{IMH}.

In this artificial setting with limited state access, the only way to observe the thermal distribution $p$ is through $P_O$ and imprecise energy measurements.
If Algorithm \ref{IMH} is given inputs $a$ that are unbiased samples from $p$,
 then it outputs labels $\gamma = (\cdots, o_1, \omega_1, o_0, \omega_0)$ that can be used to construct unbiased estimators of observables,
\begin{align} \label{classical_estimators}
 f(o_1) &\sim \mu_f = \sum_{a \in \mathcal{S}} f_O(a) p(a), \ \ \ f_O(a) = \sum_{b \in \mathcal{S}} \sum_{i \in \mathcal{O}} f(i) P_O(i,b|a), \notag \\
 \omega_0 &\sim \mu_\omega = \sum_{a \in \mathcal{S}} \int_{\mathbb{R}} d\omega \frac{e^{-[\omega-E(a)]^2/(2\sigma^2)}}{\sqrt{2 \pi} \sigma} \omega \, p(a) = \sum_{a \in \mathcal{S}} E(a) p(a),
\end{align}
 for any observable function $f:\mathcal{O} \rightarrow \mathbb{R}$.
For both estimators, limited access to the thermal state increases sampling variance from the intrinsic variance of sampling from $\mathcal{S}$,
\begin{equation}
 \sigma^2_{f, 0} = \sum_{a} [ f_O(a) - \mu_f ]^2 p(a), \ \ \
 \sigma^2_{\omega, 0} = \sum_{a \in \mathcal{S}} [ E(a) - \mu_\omega ]^2 p(a) ,
\end{equation}
 to the actual variances of sampling from $\mathcal{O}$ and $\mathbb{R}$,
\begin{align}
 \sigma^2_{f} &= \sum_{a,b \in \mathcal{S}} \sum_{i \in \mathcal{O}} [ f(i) - \mu_f ]^2 P_O(i,b|a) p(a) \notag \\
  &= \sigma^2_{f, 0} + \frac{1}{2} \sum_{a,b,c \in \mathcal{S}} \sum_{i,j \in \mathcal{O}} [ f(i) - f(j) ]^2 P_O(i,b|a) P_O(j,c|a) p(a), \notag \\
 \sigma^2_{\omega} &= \sum_{a \in \mathcal{S}} \int_{\mathbb{R}} d\omega \frac{e^{-[\omega-E(a)]^2/(2\sigma^2)}}{\sqrt{2 \pi} \sigma} (\omega - \mu_\omega)^2 \, p(a) = \sigma^2_{\omega, 0} + \sigma^2 .
\end{align}
Any bias of the input $a$ will create bias in the output estimators, which I analyze further in the next subsection for bias caused by finite values of $n_{\max}$ in Algorithm \ref{IMH}.

To compare Algorithms \ref{MH} and \ref{IMH}, I decompose $\widetilde{P}_I$ as a conditional probability function that is conditioned on the loop iteration number $n$,
\begin{align} \label{IMH_probability3}
 \widetilde{P}_I(b|a) &= \sum_{n=1}^{n_{\max}} P_I(b | n, a) [1-\lambda_n(a)] \prod_{m \in \{1, \cdots , n\} \setminus \{ n \} } \lambda_m(a) + P_T (b| a) \prod_{n=1}^{n_{\max}} \lambda_n(a),
\end{align}
 where $\lambda_n(a)$ is the probability of rejection after branch $n$ starting from state $a$ defined by
\begin{equation}
 \lambda_n(a) = 1 - \frac{\sum_{b \in \mathcal{S}} \int_{\mathcal{M}_n} d\mu(\gamma) P_I(\gamma, b | a)}{ \prod_{m \in \{1, \cdots , n\} \setminus \{ n \} } \lambda_m(a) }.
\end{equation}
The conditional probability function for each number of loop iterations can be constructed by marginalizing over outcomes associated with each iteration and applying Bayes' rule,
\begin{equation}
P_I(b|n,a) = \frac{\int_{\mathcal{M}_n} d\mu(\gamma) P_I(\gamma, b | a)}{[1 - \lambda_n(a)] \prod_{m \in \{1, \cdots , n\} \setminus \{ n \} } \lambda_m(a) },
\end{equation}
 and the terminal conditional probability function accounts for the remaining probability,
\begin{equation}
P_T(b|a) = \frac{\int_{\mathcal{M}_{n_{\max}}} d\mu(\gamma) [\widetilde{P}_I(\gamma, b | a) - P_I(\gamma, b| a)]}{ \prod_{n=1}^{n_{\max}} \lambda_n(a) }.
\end{equation}
The loop-based decomposition of $\widetilde{P}_I$ in Eq.\@ (\ref{IMH_probability3}) is similar to $P_M^{\le n}$ in Eq.\@ (\ref{acceptance_split}) for $n= n_{\max}$,
 except that $P_M^{\le n}$ returns to the input state after it rejects $n_{\max}$ states to maintain detailed balance while $\widetilde{P}_I$ halts with a terminal process $P_T$ that does not satisfy detailed balance.
Also, each branch is different for $\widetilde{P}_I$ while the first $n$ branches of $P_M^{\le n}$ are the same.
From this perspective, both algorithms appear similarly productive at generating samples from the thermal distribution.
The main difference is that Algorithm \ref{MH} gets stuck returning the same sample after repeated rejections while Algorithm \ref{IMH} gets stuck in its internal loop.
A problem specific to Algorithm \ref{IMH} is that $\lambda_n(a)$ is not bounded away from one in the $n \rightarrow \infty$ limit,
 whereas Algorithm \ref{IMH} has a constant rejection probability $\lambda(a)$ for each branch.
The consequences and mitigation of this problem are discussed in the next two subsections.

\subsection{Error analysis\label{classical_error_section}}

Because the terminal branch $P_T$ of $\widetilde{P}_I$ in Eq.\@ (\ref{IMH_probability3}) does not satisfy detailed balance,
 the stationary distribution $\tilde{p}$ of $\widetilde{P}_I$ deviates from the thermal distribution $p$.
Consequences of this deviation can be understood using standard error analysis techniques from numerical linear algebra.
The primary quantity $\tilde{\epsilon}$ of this error analysis is a bound on the change in $\tilde{p}$ caused by applying $P_I$ as measured by total variation distance,
\begin{equation} \label{define_error}
 \frac{1}{2} \sum_{a \in \mathcal{S}}\left| \sum_{b \in \mathcal{S}} P_I(a|b) \tilde{p}(b) - \tilde{p}(a) \right| \le \sum_{a \in \mathcal{S}} \tilde{p}(a) \prod_{m=1}^{n_{\max}} \lambda_m(a) = \tilde{\epsilon},
\end{equation}
 which results from $P_I$ and $\widetilde{P}_I$ being identical for their first $n_{\max}$ branches and assuming the maximum possible total variation distance of one beyond that.
With this assumption, the details of the terminal branch $P_T$ are irrelevant to the error analysis.
Similar to $\tilde{\epsilon}$, $\epsilon$ is a bound on the amount that $\widetilde{P}_I$ changes $p$ as measured by total variation distance,
\begin{equation}
 \frac{1}{2} \sum_{a \in \mathcal{S}}\left| \sum_{b \in \mathcal{S}} \widetilde{P}_I(a|b) p(b) - p(a) \right| \le \sum_{a \in \mathcal{S}} p(a) \prod_{m=1}^{n_{\max}} \lambda_m(a) = \epsilon .
\end{equation}
The retention rates $\Omega_I$ and $\widetilde{\Omega}_I$ for $P_I$ and $\widetilde{P}_I$ as defined by Eq.\@ (\ref{mix_time}) are also important to the error analysis
 as condition numbers that amplify errors beyond $\epsilon$ and $\tilde{\epsilon}$.

The total variation distance between $p$ and $\tilde{p}$ can be related to $\tilde{\epsilon}$ and $\Omega_I$ in steps as
\begin{align} \label{error_derive}
  \frac{1}{2} \sum_{a \in \mathcal{S}} | \tilde{p}(a) - p(a) | &= \frac{1}{2} \sum_{a \in \mathcal{S}} \left| \tilde{p}(a) - \sum_{b \in \mathcal{S}} P_I(a|b) \tilde{p}(b) + \sum_{b \in \mathcal{S}} P_I(a|b) [\tilde{p}(b) - p(b)] \right|  \notag \\
  &\le \tilde{\epsilon} + \frac{ \sum_{a \in \mathcal{S}} \left| \sum_{b \in \mathcal{S}} P_I(a|b) [\tilde{p}(b) - p(b)] \right| }{ 2 \sum_{a \in \mathcal{S}} | \tilde{p}(a) - p(a)| } \sum_{a \in \mathcal{S}} | \tilde{p}(a) - p(a)| \notag \\
  & \le \tilde{\epsilon} + \Omega_I \frac{1}{2} \sum_{a \in \mathcal{S}} | \tilde{p}(a) - p(a)|,
\end{align}
 where common terms are added and subtracted so that the triangle inequality can be used to isolate $\tilde{\epsilon}$ followed by
 multiplying and dividing by common terms and relaxing $\tilde{p} - p$ to the maximization in Eq.\@ (\ref{mix_time}) that isolates $\Omega_I$.
This distance is then bounded as
\begin{equation} \label{p_error}
 \frac{1}{2} \sum_{a \in \mathcal{S}} | \tilde{p}(a) - p(a) | \le \frac{\tilde{\epsilon}}{1 - \Omega_I},
\end{equation}
 and a similar derivation can be used to derive the complementary bound,
\begin{equation} \label{p_error2}
 \frac{1}{2} \sum_{a \in \mathcal{S}} | \tilde{p}(a) - p(a) | \le \frac{\epsilon}{1 - \widetilde{\Omega}_I}.
\end{equation}
Other bounds can then be constructed by further relaxing these bounds.

Error bounds are more useful when they depend on quantities that can be measured.
In this analysis, $\tilde{\epsilon}$ can be estimated from halting statistics of the loop in Algorithm \ref{IMH},
 and $\widetilde{\Omega}_I$ can be bounded from below using coarse-grained analysis as discussed in Sec.\@ \ref{classical_practical}.
This is not sufficient to evaluate strict error bounds, although approximated bounds may still serve as error estimates.
Furthermore, $\epsilon$ and $\Omega_I$ cannot be measured directly,
 and $\epsilon \approx \tilde{\epsilon}$ and $\Omega_I \approx \widetilde{\Omega}_I$ are restrictive assumptions for an error analysis.
This is motivation to relax the error bound in Eq.\@ (\ref{p_error2}) by relating $\epsilon$ and $\tilde{\epsilon}$ as
\begin{equation}
 \epsilon \le \tilde{\epsilon} + \epsilon_{\max} \frac{1}{2} \sum_{a \in \mathcal{S}} | \tilde{p}(a) - p(a) | , \ \ \ \epsilon_{\max} = \max_{a \in \mathcal{S}} \prod_{m=1}^{n_{\max}} \lambda_m(a),
\end{equation}
using H\"{o}lder's inequality.
The looser but more measureable form of Eq.\@ (\ref{p_error2}) is then
\begin{equation} \label{p_error3}
 \frac{1}{2} \sum_{a \in \mathcal{S}} | \tilde{p}(a) - p(a) | \le \frac{\tilde{\epsilon}}{\max\{0,1 - \widetilde{\Omega}_I - \epsilon_{\max}\}}.
\end{equation}
Just as with $\widetilde{\Omega}_I$, the estimation of $\epsilon_{\max}$ is limited to a biased coarse-grained analysis using observations in $\mathcal{O}$ as a proxy for hidden states in $\mathcal{S}$.

Since the stationary state of $\widetilde{P}_I$ is $\tilde{p}$ instead of $p$, the expectation values of $f(o_1)$ and $\omega_0$ from the Markov chain generated by Algorithm \ref{IMH}
 are distorted from Eq.\@ (\ref{classical_estimators}) into
\begin{equation} \label{classical_estimators2}
 f(o_1) \sim \tilde{\mu}_f = \sum_{a \in \mathcal{S}} f_O(a) \tilde{p}(a), \ \ \
 \omega_0 \sim \tilde{\mu}_\omega = \sum_{a \in \mathcal{S}} E(a) \tilde{p}(a),
\end{equation}
 with similar distortions to their sampling variances.
The error in these expectation values can then be bounded using H\"{o}lder's inequality and Eq.\@ (\ref{p_error3}) as
\begin{equation} \label{ev_error}
 | \tilde{\mu}_f - \mu_f | \le \frac{2 \tilde{\epsilon} \max_{i \in \mathcal{O}} | f(i) | }{\max\{0,1 - \widetilde{\Omega}_I - \epsilon_{\max}\}} , \ \ \
 | \tilde{\mu}_\omega - \mu_\omega | \le  \frac{2 \tilde{\epsilon} \max_{a \in \mathcal{S}} | E(a) | }{\max\{0, 1 - \widetilde{\Omega}_I - \epsilon_{\max}\}},
\end{equation}
 where the maximization of $|f_O(a)|$ for $a \in \mathcal{S}$ has been relaxed further into a maximization involving $f$ because $f_O$ is difficult to evaluate directly.
In practice, these error bounds can be useful as error estimates by using biased estimated values for the upper bounds.

\subsection{Halting analysis\label{halting_section}}

The expected run time of Algorithm \ref{IMH} is proportional to the mean of the distribution over the number of iterations needed to halt its loop.
Halting is determined by the value of $x$ at the beginning of each loop iteration, with a chance to halt if $x$ attains a new minimum value and a guaranteed halt if $x$ becomes negative.
Mathematically, $x$ is the real-valued component of a Markov chain on $\mathbb{R} \times \mathcal{O} \times \mathcal{S}$
 where each loop iteration generates an $(x, o, a)$ value that depends on the $(x, o, a)$ value of the previous iteration.
This can be interpreted as a random walk on $\mathbb{R}$ with an internal state on $\mathcal{O} \times \mathcal{S}$.
Because the halting decision does not influence the walk, the distribution of walks can be extended to infinite length beyond the actual halting point.
A detailed analysis of the halting distribution that is defined by this random walk would be complicated,
 and I focus this subsection on a special case and a simple example that provide insights into a more general analysis.
 
First, I show that the expected halting time is infinite for $\sigma > 0$ in the special case of direct state access
 with $\mathcal{O} = \mathcal{S}$, $P_C(b|i,a) = \delta_{b,i}$, $P_O(i,b|a) = \delta_{i,a} \delta_{b,a}$, and a $P$ without any forbidden transitions.
In this case, the halting statistics for each instance of Algorithm \ref{IMH} reduce to a Markov chain on $x \in \mathbb{R}$ that depends on the initial samples $i$ from $P_O$ and $o$ from $P$.
The overall expected halting time will be infinite if it is infinite for even a single allowed transition, which I show for null transitions with $i = o$.
The halting distribution for null transitions is also a complete statistical description for the simple example of an idle update, $P(o|i) = \delta_{o,i}$.
Both reduce to a discrete process with the halting condition
\begin{equation} \label{simple_halt_process}
 e^{y_n} \ge \left\{ \begin{array}{cc} u_n e^{y_0 + \Delta} , & n = 1 \\
 u_n e^{y_0 + \Delta} + (1 - u_n) \max_{m \in \{1, \cdots , n-1\} }e^{y_m}, & n \ge 2 \end{array} \right.
\end{equation}
 for step $n \ge 1$, with $u_n$ sampled from the uniform distribution over $[0,1]$ and $y_n$ sampled from the normal distribution of mean zero and variance $\Delta = \beta^2 \sigma^2$.
I evaluate the halting distribution $p_{\mathrm{halt}}$ of this process and some of its asymptotic properties in Appendix \ref{halting_distribution}.

The probability of accepting a null transition before the first loop iteration is
\begin{equation}
 p_{\mathrm{halt}}(1) = \mathrm{erfc}(\beta \sigma/2) \approx 1 - \beta \sigma / \sqrt{\pi},
\end{equation}
 which guarantees acceptance in the $\beta \sigma \rightarrow 0$ limit and has a linear relationship with $\sigma$ for $\beta \sigma \ll 1$.
However, the halting distribution has a fat tail that is asymptotically bounded as
\begin{equation}
  p_{\mathrm{halt}}(n) = \omega(1/n^2)
\end{equation}
 for loop iteration $n$.
This increases the $n_{\max}$ sensitivity of the truncation error in Eq.\@ (\ref{define_error}),
\begin{equation}
 \tilde{\epsilon} \approx \frac{\beta \sigma e^{\beta \sigma \sqrt{2 \log n_{\max}}}}{\sqrt{2 \pi} n_{\max} \log n_{\max}} = \omega(1/n_{\max})
\end{equation}
 and causes a divergence in the expected halting time,
\begin{equation}
 \tilde{n}_{\mathrm{halt}} = n_{\max} - \sum_{n=1}^{n_{\max}} (n_{\max} - n) p_{\mathrm{halt}}(n) \approx e^{\beta \sigma \sqrt{2 \log n_{\max}} - \beta^2 \sigma^2 / 2 },
\end{equation}
 in the $n_{\max} \rightarrow \infty$ limit.
The cost-accuracy relationship between this $\tilde{n}_{\mathrm{halt}}$ and $\tilde{\epsilon}$ is
\begin{equation} \label{cost_accuracy}
  \tilde{n}_{\mathrm{halt}} \approx e^{\beta \sigma \sqrt{2 \log (1 / \tilde{\epsilon})}}
\end{equation}
 for $\tilde{\epsilon} \ll \beta \sigma \ll 1$, which is worse than the typical $O(\log(1/\tilde{\epsilon}))$ cost of numerical errors but better than the $O(1/\tilde{\epsilon}^2)$ cost of sampling errors.
However, the cause of this cost multiplier is the stalling of a repeat-until-success loop that is not performing any useful computation
 beyond waiting for a success criterion that becomes less likely with each loop iteration.
If $\sigma$ can be modified, then the growth of $\tilde{n}_{\mathrm{halt}}$ can be mitigated by $\sigma \propto 1/(\beta \sqrt{\log(1/\tilde{\epsilon})})$.
 
Beyond the case of null transitions, transitions between distinct states $a$ and $b$ change the discrete halting process in Eq.\@ (\ref{simple_halt_process})
 by shifting the mean of $y_n$ for odd $n$ from zero to $\beta[ E(a) - E(b) ]$.
For $E(b) \gg E(a)$, the expected halting time will double because halting will almost never occur for odd $n$ and be unchanged for even $n$.
For $E(b) \ll E(a)$, there will still be a fat tail in the halting distribution and a divergence in the expected halting time,
 but the weight of the tail will be suppressed by a Boltzmann weight.
Beyond these small effects, the halting distribution for a general $P$ will be qualitatively the same as the idle update when direct state access is used in Algorithm \ref{IMH}.
 
The halting process is substantially more complicated when direct state access is not used in Algorithm \ref{IMH}.
The $x_n$ values from Eq.\@ (\ref{simple_accept}) for $n \ge 2$ have the more general form
\begin{equation}
 x_n = x_0 - e^{-\beta \omega_n} P(o_n | o_{n-1}) + \sum_{r=1}^{n-1} e^{-\beta \omega_r} [ P(o_r | o_{r+1}) - P(o_r | o_{r-1}) ]
\end{equation}
 in which $x_n - x_0$ are no longer statistically independent random variables.
Superficially, the sum in $x_n - x_0$ looks like the accumulated value of a random walk on $\mathbb{R}$, albeit with a non-trivial internal state.
I would expect this random walk to have a bias proportional to $P(o_n | o_{n+1}) - P(o_n | o_{n-1})$, which should have a zero expectation value because the Markov chain
 of $o_n$ is generated by the symmetric process $P_S$ from Eq.\@ (\ref{classical_spam}).
An unbiased random walk should then only add more variance to the halting process over the baseline variance from the uncertain energy measurements.
The qualitative behavior of the halting process should remain the same as the simple example, with a larger combined variance.
Details of how these variances combine are complicated and beyond the scope of this paper.

\section{Quantum algorithm}

Finally, I adapt Algorithm \ref{IMH} into a quantum algorithm that stabilizes a quantum thermal state.
I replace its classical SPAM and energy operations with quantum operations, then I describe it as a quantum operation instead of as a conditional probability function.
The quantum operation describing the quantum algorithm satisfies quantum detailed balance, which guarantees the stability of the quantum thermal state.

The classical SPAM operations in Sec.\@ \ref{imprecise_section} were split into a measurement operation $P_O$ and a control operation $P_C$
 acting on sets of states $\mathcal{S}$ and observations $\mathcal{O}$.
In a quantum setting, classical states on $\mathcal{S}$ are replaced with a set of normalized quantum states on a Hilbert space, denoted by $\mathbb{H}(\mathcal{S})$.
This Hilbert space is spanned by basis states $|a\rangle \in \mathbb{H}(\mathcal{S})$ for $a \in \mathcal{S}$ such that $\langle a | b \rangle = \delta_{a,b}$ in bra-ket notation.
Quantum operations such as control and measurement are defined by a positive operator-valued measure (POVM), which is a set of linear Kraus operators on $\mathbb{H}(\mathcal{S})$,
 $K(x)$ for $x \in \mathcal{X}$, and a measure $\nu$ on $\mathcal{X}$ satisfying
\begin{equation} \label{K_norm}
 \int_{\mathcal{X}} d\nu(x) K^\dag(x) K(x) = I,
\end{equation} 
 where $I$ is the identity operator and $X^\dag$ is the conjugate transpose of $X$.
A POVM is a stochastic map on pure quantum states, 
$|\psi\rangle \in \mathbb{H}(\mathcal{S})$, with the deterministic state map
\begin{equation} \label{K_collapse}
 |\psi\rangle \mapsto \frac{K(x)|\psi\rangle}{\sqrt{\langle \psi | K^\dag(x) K(x)|\psi\rangle}}
\end{equation}
 conditioned on a value of $x$ sampled from the probability measure $\langle \psi | K^\dag(x) K(x)|\psi\rangle$.
It is equivalently a trace-preserving linear map on quantum density matrices $\rho$ in $\mathbb{H}(\mathcal{S})$,
\begin{equation}
  \rho \mapsto  \int_{\mathcal{X}} d\nu(x) K(x) \rho K^\dag(x).
\end{equation}
A quantum measurement operation is a POVM on the set $\mathcal{O}$ with the counting measure and Kraus operators $K_O$.
Here, I consider unitary quantum control operations, which are a set of POVMs indexed by $o \in \mathcal{O}$, each on a trivial set with one element and a unitary Kraus operator $U_C(o)$.
The quantum version of the symmetry condition in Eq.\@ (\ref{classical_spam}) is
\begin{equation} \label{quantum_spam}
 U_C(o) K_O(i)  =  K_O^\dag(o) U_C^\dag(i)
\end{equation}
 for all $i,o \in \mathcal{O}$.
A typical example of quantum SPAM operations is
\begin{align}
 U_C(o) &= I + \sum_{n=1}^{|\mathcal{S}|/|\mathcal{O}|} \left( |\kappa_n(o) \rangle \langle \kappa_n(j) | + |\kappa_n(j) \rangle \langle \kappa_n(o) | 
  - |\kappa_n(o) \rangle \langle \kappa_n(o) | - |\kappa_n(j) \rangle \langle \kappa_n(j) | \right), \notag \\
 K_O(i) &= \sum_{n=1}^{|\mathcal{S}|/|\mathcal{O}|} |\kappa_n(j) \rangle \langle \kappa_n(i) |,
\end{align}
 defined by an orthonormal basis of $\mathbb{H}(\mathcal{S})$, $|\kappa_n(o)\rangle$ for $o \in \mathcal{O}$ and $n \in \{1, \cdots, |\mathcal{S}|/|\mathcal{O}| \}$,
 and a $j \in \mathcal{O}$ that corresponds to a post-measurement subspace of this basis.

The quantum generalization of the energy measurement operation in Sec.\@ \ref{imprecise_section} is also a POVM.
It is an ideal, Gaussian-filtered form of QPE \cite{qpe} defined by Kraus operators
\begin{equation} \label{energy_kraus}
 K_E(\omega) = \frac{\exp[-(\omega - H)^2/(4\sigma^2)]}{\sqrt[4]{2\pi \sigma^2}}
\end{equation}
 for a measured energy $\omega \in \mathbb{R}$, an energy uncertainty $\sigma \in \mathbb{R}_{\ge 0}$,
  and a Hamiltonian $H$ that is a Hermitian operator on $\mathbb{H}(\mathcal{S})$ with a spectral decomposition labelled by $a \in \mathcal{S}$, 
\begin{equation}
 H |a\rangle = E(a) | a \rangle ,
\end{equation}
 which is a quantum generalization of a classical energy function, $E:\mathcal{S} \rightarrow \mathbb{R}$.
The measure for $\mathbb{R}$ here is the Lebesgue measure, corresponding to standard integration.
The classical energy measurement is recovered when $K_E(\omega)$ is applied to $|a\rangle$,
 which preserves $|a\rangle$ and samples $\omega$ from the normal distribution with mean $E(a)$ and variance $\sigma^2$.

\begin{algorithm}[t]
\caption{Quantum Metropolis--Hastings update \label{QMH}}
\begin{algorithmic}[1]
\Require{ $n_{\max} \in \mathbb{N}$, quantum state $|\psi\rangle \in \mathbb{H}(\mathcal{S})$, reciprocal temperature $\beta \in \mathbb{R}_{\ge 0}$, energy uncertainty $\sigma \in \mathbb{R}_{\ge 0}$,
 Kraus operators $K_E(\omega)$ for $\omega \in \mathbb{R}$ defined by Eq.\@ (\ref{energy_kraus}),
 unitary operators $U_C(o)$ and Kraus operators $K_O(o)$ for $o \in \mathcal{O}$ satisfying Eq.\@ (\ref{quantum_spam})},
 probability distribution $P$ over $\mathcal{O}$ conditioned on $\mathcal{O}$
\Ensure{$|\psi\rangle$ is replaced by $\widetilde{K}_I(\gamma) |\psi\rangle$ defined by Eqs.\@ (\ref{R_recursion}), (\ref{IMH_probability2}), (\ref{generic_accept}), (\ref{IMH_quantum}), and (\ref{quantum_sym})}
\State $(\omega_0, |\psi\rangle) \gets$ measurement from $K_E(\cdot) |\psi\rangle$
\State $(i, |\psi\rangle) \gets$ measurement from $K_O(\cdot) |\psi\rangle$
\State $o \gets$ sample from $P(\cdot | i)$
\State $|\psi\rangle \gets U_C(o) |\psi\rangle$
\State $x_{\min} \gets \exp(-\beta \omega_0 + \beta^2 \sigma^2) P(o|i)$
\State $(\omega, |\psi\rangle) \gets$ measurement from $K_E(\cdot) |\psi\rangle$
\State $x \gets x_{\min} - \exp(- \beta \omega) P(i|o)$
\State $u \gets \textrm{sample from the uniform distribution over} \ [0,1]$
\State $ \gamma \gets (i, \omega, o, \omega_0)$
\State $(o, n) \gets (i, 1)$
\While{$x > x_{\min} u \ \mathrm{and} \ n < n_{\max}$}
\State $(i, |\psi\rangle) \gets \textrm{measurement from} \ K_O(\cdot) |\psi\rangle$
\State $|\psi\rangle \gets U_C |\psi\rangle$
\State $x_{\min} \gets \min\{ x_{\min} , x \}$
\State $x \gets x + \exp(- \beta \omega) P(o|i)$
\State $(\omega, |\psi\rangle) \gets$ measurement from $K_E(\cdot) |\psi\rangle$
\State $x \gets x - \exp(- \beta \omega) P(i|o) $
\State $u \gets \textrm{sample from the uniform distribution over} \ [0,1]$
\State $ \gamma \gets (i, \omega, \gamma)$
\State $(o, n) \gets (i, n+1)$
\EndWhile
\end{algorithmic}
\end{algorithm}

Now I can convert Algorithm \ref{IMH} into a quantum algorithm by replacing $P_O$, $P_C$, and $E$ with their corresponding quantum operations based on $K_O$, $U_C$, and $K_E$.
This results in Algorithm \ref{QMH}, which extends the classical pseudocode notation to include quantum states.
While there are proposals for quantum pseudocode notation \cite{quantum_pseudocode}, there are conflicts with classical notation and no consensus.
Classical pseudocode also has no rigid conventions, but variables are usually passed to functions as copies, and the value of a variable only changes by explicit assignment.
Since an unknown quantum state cannot be copied, this is not a practical convention for quantum pseudocode.
Instead, I assume that any quantum variable passed to a function or otherwise operated on is altered by the computation.
For consistency with classical notation, a quantum variable cannot be reused in computation until it is explicitly assigned a new value,
 often as the outcome of the same computation that changed its value.
This labeling of quantum variables on both sides of an assignment is similar in concept to the use of wires as input and output labels in quantum circuits.

A sequence of quantum states generated by repeatedly running Algorithm \ref{QMH} and using the quantum output of each run as the quantum input of the next run
 defines a quantum Markov chain that is intended to stabilize a thermal density matrix,
\begin{equation} \label{quantum_thermal}
 \rho = \sum_{a \in \mathcal{S}} \frac{\exp[-\beta E(a)]}{Z} | a\rangle \langle a |,
\end{equation}
 analogous to the thermal distribution in Eq.\@ (\ref{thermal}).
As with the quantum SPAM and energy operations, I describe Algorithm \ref{QMH} as a POVM with Kraus operators $\widetilde{K}_I(\gamma)$ for $\gamma \in \mathcal{M}$,
 with the same set $\mathcal{M}$ and measure $\mu$ as used in Eq.\@ (\ref{IMH_probability}) from Sec.\@ \ref{imprecise_section}.
I also consider an ideal POVM with Kraus operators $K_I(\gamma)$, which are the $n_{\max} \rightarrow \infty$ limit of $\widetilde{K}_I(\gamma)$.
If the quantum Markov chain is initialized to a known state $|\psi_0\rangle$, then each state in the chain is
\begin{equation}
 |\psi_n\rangle = \frac{\widetilde{K}_I(\gamma_n)|\psi_{n-1}\rangle}{\sqrt{\langle \psi_{n-1} | \widetilde{K}_I^\dag(\gamma_n) \widetilde{K}_I(\gamma_n)|\psi_{n-1}\rangle}},
\end{equation}
 where $\gamma_n$ is the classical outcome of the quantum operation in step $n$ of the chain.
Such a quantum Markov chain decomposes into a deterministic quantum process conditioned on the random outcome of a statistical classical process.

I decompose the quantum process like the classical process in Eqs.\@ (\ref{IMH_probability}) and (\ref{classical_split}) as
\begin{equation} \label{IMH_quantum}
 \langle b | \widetilde{K}_I(\gamma_n) | a \rangle = \sum_{\mathbf{s} \in \mathcal{S}^{n-1}} \kappa(\gamma_n, b, \mathbf{s}, a) \sqrt{\widetilde{P}_I^{\mathrm{dec}}(\gamma_n)},
\end{equation}
 with an equivalent decomposition of $K_I(\gamma)$ that replaces $\widetilde{P}_I^{\mathrm{dec}}$ by $P_I^{\mathrm{dec}}$.
This combines the classical decision components in Eqs.\@ (\ref{R_recursion}) and (\ref{IMH_probability2}) with a quantum symmetric component
\begin{equation} \label{quantum_sym}
 \kappa(\gamma_n, a_n, \cdots , a_0) = \left( \prod_{m=0}^n  \frac{e^{-[\omega_m-E(a_m)]^2/(4\sigma^2)}}{\sqrt[4]{2 \pi \sigma^2}} \right) \prod_{m=0}^{n-1} \langle a_{m+1} | U_C(o_m) K_O(o_{m+1}) | a_m \rangle
\end{equation}
 for $\gamma_n = (o_n, \omega_n, \cdots, o_0, \omega_0) \in \mathcal{M}_n$,
 which corresponds to a sequence of virtual quantum transitions between stationary states instead of real but unknown classical transitions.

Similar to Sec.\@ \ref{imprecise_section}, I will replace the standard quantum detailed balance condition \cite{quantum_metropolis}
 with a finer balance condition on $K_I$ between subsets of classical outcomes as a sufficient condition for $\rho$ to be a stationary state of the ideal quantum process,
\begin{equation} \label{quantum_stationary}
 \int_{\mathcal{M}} d\mu(\gamma) K_I(\gamma) \rho K^\dag_I(\gamma) = \rho .
\end{equation}
A simple example of this type of balance is the condition
\begin{equation} \label{quantum_balance}
 \langle a | K_I(\gamma) | b \rangle e^{-\beta E(b)/2} = \langle a | K_I^\dag( \alpha(\gamma) ) | b \rangle e^{-\beta E(a)/2}
\end{equation}
 for all $a,b \in \mathcal{S}$ and $\gamma \in \mathcal{M}$, where $\alpha$ is a measure-preserving automorphism on $\mathcal{M}$.
This is a sufficient condition for Eq.\@ (\ref{quantum_stationary}), which can be shown in four steps,
\begin{align} \label{quantum_balance_derive}
 \int_{\mathcal{M}} d\mu(\gamma) K_I(\gamma) \rho K^\dag_I(\gamma) &=
 \frac{1}{Z} \sum_{a,b,c \in \mathcal{S}}  \int_{\mathcal{M}} d\mu(\gamma) e^{-\beta E(b)} |a \rangle \langle a| K_I(\gamma) | b \rangle \langle b | K^\dag_I(\gamma) | c \rangle \langle c | \notag \\
 &= \frac{1}{Z} \sum_{a,b,c \in \mathcal{S}}  \int_{\mathcal{M}} d\mu(\gamma) e^{-\beta [ E(a) + E(c) ] /2} |a \rangle \langle a| K^\dag_I(\gamma) | b \rangle \langle b | K_I(\gamma) | c \rangle \langle c | 
\notag \\
 &= \int_{\mathcal{M}} d\mu(\gamma) \sqrt{\rho} K^\dag_I(\gamma) K_I(\gamma) \sqrt{\rho} = \rho,
\end{align}
 where $\rho$ and $K_I$ are first projected into the basis $|a\rangle$ for $a \in \mathcal{S}$, then Eq.\@ (\ref{quantum_balance}) is applied to both $K_I$ with $\alpha$ absorbed into the measure,
 then the projection is removed, and finally $K_I$ is removed using Eq.\@ (\ref{K_norm}).
Unfortunately, Eq.\@ (\ref{quantum_balance}) is too strict of a balance condition for $K_I$ to satisfy because $E(a)$ cannot be measured and used in acceptance probabilities.

To accommodate energy uncertainty, I loosen the balance condition from Eq.\@ (\ref{quantum_balance}) to
\begin{align} \label{quantum_balance2}
 & \int_{\mathbb{R}^2} d\omega \, d\omega' \langle a | K_I(o, \omega, \gamma, o', \omega') | b \rangle \langle b | K_I^\dag(o, \omega, \gamma, o', \omega') | c \rangle e^{-\beta E(b)} = \notag \\
 & \int_{\mathbb{R}^2} d\omega \, d\omega' \langle a | K_I^\dag( o', \omega', \overline{\gamma}, o, \omega ) | b \rangle \langle b | K_I( o', \omega', \overline{\gamma}, o, \omega ) | c \rangle  e^{-\beta [E(a) + E(c)]/2}
\end{align}
  for all $a, b, c \in \mathcal{S}$, $o,o' \in \mathcal{O}$, and $\gamma \in \mathcal{M}$ and the specific automorphism $\alpha(\gamma) = \overline{\gamma}$ used in Eq.\@ (\ref{branch_balance3}).
Eq.\@ (\ref{quantum_balance2}) can then be used instead of Eq.\@ (\ref{quantum_balance}) to derive Eq.\@ (\ref{quantum_balance_derive}).
In the presence of these $\omega$ and $\omega'$ integrals, I use Eq.\@ (\ref{gaussian_identity}) to switch Boltzmann weights from $E$ to $\omega$ and $\omega'$ in Eq.\@ (\ref{quantum_balance2}) after regrouping products of Gaussians as
\begin{equation}
 e^{- (\omega - E)^2/(4\sigma^2)} e^{- (\omega - E')^2/(4\sigma^2)} = e^{- [\omega - (E + E')/2]^2/(2\sigma^2)} e^{-(E - E')^2/(8\sigma^2)} .
\end{equation}
Following this switch, I isolate a simple sufficient condition for Eq.\@ (\ref{quantum_balance2}) that applies strict detailed balance to time-reversed pairs of Kraus operators,
\begin{equation} \label{quantum_balance3}
 K_I(\gamma + \beta \sigma^2) e^{-\beta \underline{\omega}_0(\gamma)/2} = K_I^\dag(\overline{\gamma} + \beta \sigma^2) e^{-\beta \underline{\omega}_0(\overline{\gamma})/2}.
\end{equation}
Finally, I use Eqs.\@ (\ref{quantum_spam}) and (\ref{IMH_quantum}) to remove a common factor of $\kappa$ from both sides of this equation
 to reduce it to an equivalent of Eq.\@ (\ref{branch_balance4}) that is again satisfied by Eq.\@ (\ref{generic_accept}).

If the quantum input $|\psi\rangle$ to Algorithm \ref{QMH} is an unbiased estimator of the thermal state, $|\psi\rangle \langle \psi | \sim \rho$,
 then it outputs unbiased estimators of thermal expectation values
\begin{align} \label{quantum_estimators}
 f(o_1) &\sim \mu_f = \mathrm{tr}(F \rho), \ \ \ F = \sum_{i \in \mathcal{O}} f(i) K_O^\dag(i) K_O(i), \notag \\
 \omega_0 &\sim \mu_\omega = \int_{\mathbb{R}} d\omega \, \omega \, \mathrm{tr}[ K_E(\omega)^2 \rho] = \mathrm{tr}( H \rho ),
\end{align}
 for any observable function $f:\mathcal{O} \rightarrow \mathbb{R}$ from the POVM outcome $\gamma = (\cdots, o_1, \omega_1, o_0, \omega_0)$.
Algorithm \ref{QMH} measures $F$ and $H$ indirectly, which increases sampling variance from the intrinsic variance of measuring in their eigenbasis,
\begin{equation}
 \sigma^2_{f, 0} = \mathrm{tr} [ (F - \mu_f  I)^2 \rho ], \ \ \
 \sigma^2_{\omega, 0} = \mathrm{tr} [ ( H - \mu_\omega I)^2 \rho ] ,
\end{equation}
  to the actual variances of measuring $K_O$ and $K_E$,
\begin{align}
 \sigma^2_{f} &= \sum_{i \in \mathcal{O}} [f(i) - \mu_f]^2 \mathrm{tr}[K_O^\dag(i) K_O(i) \rho] \notag \\
 &= \sigma^2_{f, 0} + \frac{1}{2} \sum_{i,j \in \mathcal{O}} [f(i) - f(j)]^2 \mathrm{tr}[ K_O^\dag(i) K_O(i) K_O^\dag(j) K_O(j) \rho ]  , \notag \\
 \sigma^2_{\omega} &= \int_{\mathbb{R}} d\omega (\omega - \mu_\omega)^2 \mathrm{tr}[ K_E(\omega)^2 \rho] = \sigma^2_{\omega, 0} + \sigma^2 .
\end{align}
Finite values of $n_{\max}$ in Algorithm \ref{QMH} will bias the stationary state $\tilde{\rho}$ of $\widetilde{K}_I$ and bias these estimators,
 which I analyze in the next subsection by adapting results from Sec.\@ \ref{classical_error_section}.

\subsection{Quantum operation analysis}

Much of the analysis in Secs.\@ \ref{classical_section} and \ref{classical_error_section} can be adapted to Algorithm \ref{QMH}
 by switching from classical to quantum primitives and adjusting derivations as needed.
Here, I describe the quantum output of Algorithm \ref{QMH} with a completely positive, trace-preserving (CPTP) map,
\begin{equation}
 \widetilde{Q}_I(\rho_0) = \int_{\mathcal{M}} d\mu(\gamma) \widetilde{K}_I(\gamma) \rho_0 \widetilde{K}_I^\dag(\gamma)
\end{equation}
 for a density matrix $\rho_0$ describing the statistics of the quantum input.
I describe the ideal algorithm with Kraus operators $K_I$ similarly as a CPTP map $Q_I$.
I also split these maps by branch as indicated by the output $n$ into sums of two trace-reducing maps each,
\begin{equation}
 \widetilde{Q}_I = Q_I^{n \le n_{\max}} + Q_T^{n_{\max}}, \ \ \ Q_I = Q_I^{n \le n_{\max}} + Q_I^{n > n_{\max}},
\end{equation}
 with the same map $Q_I^{n \le n_{\max}}$ for outputs $n \le n_{\max}$ and a terminal map $Q_T^{n_{\max}}$ that does not satisfy quantum detailed balance.
I continue to use the total variation distance in the analysis, which corresponds to the trace distance, $\| \rho - \rho' \|_1 / 2$ for a pair of density matrices $\rho$ and $\rho'$,
 and uses the trace norm, $\| X \|_1 = \mathrm{tr}(\sqrt{X^\dag X})$.

The concept of a maximum retention rate translates directly from Sec.\@ \ref{classical_section}.
A quantum Markov chain initialized to $\rho_0$ and generated by $Q_I$ has the reduced density matrix
\begin{equation}
 \rho_{n} = Q_I(\rho_{n-1})
\end{equation}
 for step $n$ of the chain.
The distance from the stationary state $\rho$ of $Q_I$ can be bounded as
\begin{equation}
 \frac{1}{2} \| \rho_n - \rho \|_1 \le \Omega_I^n \frac{1}{2} \| \rho_0 - \rho \|_1
\end{equation}
 for $n \ge 1$, and the maximum retention rate $\Omega_I$ has the variational definition
\begin{equation}
 \Omega_I = \max_{X} \frac{ \| Q_I(X) \|_1}{ \| X \|_1} \ \ \ \mathrm{subject \ to} \ \ \ \mathrm{tr}(X) = 0,
\end{equation}
 which is the second largest eigenvalue of $Q_I$ as optimized over the subspace orthogonal to the left eigenvector $I$ corresponding to the eigenvalue one and the right eigenvector $\rho$.
A similar retention rate $\widetilde{\Omega}_I$ can be defined for $\widetilde{Q}_I$ with respect to its stationary state $\tilde{\rho}$.

The error analysis from Sec.\@ \ref{classical_error_section} also directly translates.
The cross stability of $\rho$ and $\tilde{\rho}$ with respect to $Q_I$ and $\widetilde{Q}_I$ can be bounded as
\begin{align}
 \frac{1}{2} \| \widetilde{Q}_I(\rho) - \rho \|_1 \le 1 - \mathrm{tr}[Q_I^{n \le n_{\max}} (\rho)] &= \epsilon, \notag \\
 \frac{1}{2} \| Q_I(\tilde{\rho}) - \tilde{\rho} \|_1 \le 1 - \mathrm{tr}[Q_I^{n \le n_{\max}} (\tilde{\rho})] &= \tilde{\epsilon},
\end{align}
 and $\tilde{\epsilon}$ can be estimated from the halting statistics of Algorithm \ref{QMH}.
There are two simple bounds on the distance between $\rho$ and $\tilde{\rho}$ with a derivation analogous to Eq.\@ (\ref{error_derive}),
\begin{equation}
 \frac{1}{2} \| \rho' - \rho \|_1 \le \frac{\epsilon}{1 - \widetilde{\Omega}_I} , \ \ \ \frac{1}{2} \| \rho' - \rho \|_1 \le \frac{\tilde{\epsilon}}{1 - \Omega_I},
\end{equation}
 which again can be loosened to a more directly observable bound with the same steps,
\begin{equation}
 \frac{1}{2} \| \rho' - \rho \|_1 \le \frac{\tilde{\epsilon}}{\max \{0, 1 - \widetilde{\Omega}_I - \epsilon_{\max} \} },
\end{equation}
 for a maximum error parameter that is now optimized over quantum states,
\begin{equation}
 \epsilon_{\max} = 1 - \min_{ |\psi\rangle \in \mathbb{H}(\mathcal{S}) } \mathrm{tr}[Q_I^{n \le n_{\max}} (|\psi\rangle \langle \psi |)] .
\end{equation}
The actual expectation values that Algorithm \ref{QMH} estimates,
\begin{equation} \label{quantum_estimators2}
 f(o_1) \sim \tilde{\mu}_f = \mathrm{tr}(F \tilde{\rho}), \ \ \ \omega_0 \sim \tilde{\mu}_\omega = \mathrm{tr}( H \tilde{\rho} ),
\end{equation}
 have error bounds that are equivalent to the classical versions in Eq.\@ (\ref{ev_error}).

The decorrelation of a classical Markov chain is studied by combining the conditional and stationary distributions on $\mathcal{S}$ into distributions on $\mathcal{S}^2$ in Eq.\@ (\ref{mix_state}).
However, this does not directly generalize to the quantum case because $Q_I$ and $\rho$ do not directly correspond to a density matrix on $\mathbb{H}(\mathcal{S}^2)$.
Instead, I define an entangling operation in the $|a\rangle$ basis,
\begin{equation}
 Q_E(\rho_0) = \sum_{a,b\in \mathcal{S}} |a\rangle \langle a | \rho_0 |b\rangle \langle b | \otimes |a\rangle \langle b | ,
\end{equation}
 that maps from $\mathbb{H}(\mathcal{S})$ to $\mathbb{H}(\mathcal{S}^2)$, and then use it to define pair density matrices,
\begin{align}
 \rho_{0}^{\mathrm{pair}} &= Q_E(\rho) = \sum_{a \in \mathcal{S}} p(a) |a\rangle \langle a | \otimes |a\rangle \langle a |, \notag \\
 \rho_{n}^{\mathrm{pair}} &= (Q_I \otimes I) (\rho_{n-1}^{\mathrm{pair}}) = \sum_{a \in \mathcal{S}} p(a) \rho^{\mathrm{mix}}_n(a) \otimes |a\rangle \langle a |, \label{quantum_pair}
\end{align}
 where $I$ here acts as the identity superoperator on $\mathbb{H}(\mathcal{S})$ and $\rho_n^{\mathrm{mix}}(a)$ is the density matrix resulting from $n$ applications of $Q_I$ to $|a\rangle \langle a|$.
Eq.\@ (\ref{mix_error}) then generalizes to
\begin{equation} \label{quantum_mix_error}
  \frac{1}{2} \| \rho_{n}^{\mathrm{pair}} - \rho \otimes \rho \|_1 \le \Omega_I^n \left(1 - \sum_{a \in \mathcal{S}} p(a)^2 \right)
\end{equation}
 for $n \ge 1$.
The quantum proof of Eq.\@ (\ref{quantum_mix_error}) follows from the steps
\begin{align} \label{quantum_derive}
\frac{\| (Q_I \otimes I) (\rho_{n-1}^{\mathrm{pair}}) - \rho \otimes \rho \|_1}{\| \rho_{n-1}^{\mathrm{pair}} - \rho \otimes \rho \|_1} &=
\frac{ \| \sum_{a \in \mathcal{S}} p(a) Q_I (\rho_{n-1}^{\mathrm{mix}}(a) - \rho ) \otimes | a \rangle \langle a | \|_1}{\| \sum_{a \in \mathcal{S}} p(a) [ \rho_{n-1}^{\mathrm{mix}}(a) - \rho ] \otimes | a \rangle \langle a | \|_1} \notag \\
 &= \frac{\sum_{a \in \mathcal{S}} p(a) \| Q_I (\rho_{n-1}^{\mathrm{mix}}(a) - \rho) \|_1}{\sum_{a \in \mathcal{S}} p(a) \| \rho_{n-1}^{\mathrm{mix}}(a) - \rho \|_1} \notag \\
 &\le \sum_{a \in \mathcal{S}} p(a) \frac{\| Q_I (\rho_{n-1}^{\mathrm{mix}}(a) - \rho) \|_1}{\| \rho_{n-1}^{\mathrm{mix}}(a) - \rho \|_1} = \Omega_I,
\end{align}
 by isolating trace norms on the first $\mathbb{H}(\mathcal{S})$ subsystem since the density matrix is diagonal on the second subsystem
 and then using Chebyshev's sum inequality as in Eq.\@ (\ref{classical_pair_proof}).

The analysis in this subsection is not useful in practice without a method to estimate $\widetilde{\Omega}_I$ and $\epsilon_{\max}$.
While $\tilde{\epsilon}$ can be estimated directly from observed data without bias, $\widetilde{\Omega}_I$ and $\epsilon_{\max}$ cannot.
The projective coarse graining from Sec.\@ \ref{classical_section} cannot be applied to a quantum Markov chain in general,
 but the conditional dependence of observations along a quantum Markov chain can still be estimated without assuming Markovian structure.
The decay of observed correlations can enable biased estimates of $\widetilde{\Omega}_I$, and conditioning halting statistics on prior observations can enable biased estimates of $\epsilon_{\max}$.

\subsection{Cost analysis}

The effective cost of generating independent samples from a quantum thermal state using Algorithm \ref{QMH} is a product of the mixing time of the quantum Markov chain,
 the expected halting time of the repeat-until-success loop, and the average cost of each sequence of $K_O$, $U_C$, and $K_E$ operations.
Here, I provide a simple estimate of these costs, which might be refined by accounting for more implementation and problem-specific details.

I consider a mixing time $n_{\mathrm{mix}}$ defined as the number of sequential, dependent samples from the Markov chain that are needed
 to achieve the same amount of variance reduction as one independent sample in estimating expectation values.
While a precise $n_{\mathrm{mix}}$ depends on a measured Hermitian operator $X$ on $\mathbb{H}(\mathcal{S})$, an upper bound depends only on $\Omega_I$,
\begin{equation}
 n_{\mathrm{mix}} = 1 + 2 \sum_{n=1}^{\infty} \frac{\mathrm{tr}[(X \otimes X) (\rho_n^\mathrm{pair} - \rho \otimes \rho)]}{\mathrm{tr}(X^2 \rho) - \mathrm{tr}(X \rho)^2} \le 1 +  2 \sum_{n=1}^{\infty} \Omega_I^n = \frac{1 + \Omega_I}{1 - \Omega_I},
\end{equation}
 for $\rho_n^\mathrm{pair}$ defined in Eq.\@ (\ref{quantum_pair}).
This bound results from $\mathrm{tr}[(X \otimes X) \rho_0^\mathrm{pair}] \le \mathrm{tr}(X^2 \rho)$
 and the bounding of $\mathrm{tr}[(X \otimes X) (\rho_n^\mathrm{pair} - \rho \otimes \rho)]$ ratios by $\Omega_I$ after rearranging terms as
\begin{equation}
 \frac{\mathrm{tr}[(X \otimes X) (\rho_n^\mathrm{pair} - \rho \otimes \rho)]}{\mathrm{tr}[(X \otimes X) (\rho_{n-1}^\mathrm{pair} - \rho \otimes \rho)]} = \frac{\mathrm{tr}[X Q_I(Y)] }{ \mathrm{tr}(X Y)},
  \ \ \ Y = \sum_{a \in \mathcal{S}} p(a) \langle a | X | a \rangle [\rho_{n-1}^{\mathrm{mix}}(a) - \rho],
\end{equation}
 which is another variational characterization of $\Omega_I$ by maximizing over $X$ and $Y$ subject to $\mathrm{tr}(Y) = 0$.
The mixing time $\tilde{n}_{\mathrm{mix}}$ of $\widetilde{Q}_I$ is similarly determined by $\widetilde{\Omega}_I$,
 although I would expect that $\tilde{n}_{\mathrm{mix}} \approx n_{\mathrm{mix}}$ when the truncation error $\tilde{\epsilon}$ is small.

The expected halting time of a quantum idle update, $K_O = U_C = I$, is equivalent to the classical example discussed in Sec.\@ \ref{halting_section}.
However, an idle update cannot mix and will correspond to $\tilde{n}_{\mathrm{mix}} = \infty$.
In the absence of a more detailed halting analysis, I will assume a heuristic model of halting time that substitutes $\sqrt{\sigma^2 + \sigma_0^2}$ for $\sigma$ in Eq.\@ (\ref{cost_accuracy}),
 where $\sigma_0$ is a baseline variance in the halting process introduced by SPAM uncertainty.
Thus, a finite $\tilde{n}_{\mathrm{mix}}$ necessitates a nonzero $\sigma_0$, and decreasing $\tilde{n}_{\mathrm{mix}}$ may require increasing $\sigma_0$.

I assume that $K_O$ and $U_C$ have a negligible cost and implement $K_E$ by measuring an auxiliary quantum variable $\omega$ after controlled Hamiltonian time evolution and a quantum Fourier transform,
 which has the exact pre-measurement form for an input $|\psi\rangle$,
\begin{equation}
  \int_{\mathbb{R}} d\omega \frac{e^{-(\omega - H)^2/(4\sigma^2)}}{\sqrt[4]{2\pi \sigma^2}} | \psi \rangle \otimes | \omega \rangle = 
  \int_{\mathbb{R}} d\omega \frac{e^{-i \omega t}}{\sqrt{2\pi}} \left[ \int_{\mathbb{R}} dt  \, e^{i H t} | \psi \rangle \otimes \left( \sqrt[4]{2\sigma^2/\pi} e^{-\sigma^2 t^2} | t \rangle \right ) \right].
\end{equation}
I limit the cost of $K_E$ by limiting Hamiltonian evolution to a time interval $[-t_{\max}, t_{\max}]$ for $t_{\max} = \sqrt{\log(1/\tilde{\epsilon})/2}/\sigma$,
 which introduces errors with probability $\approx \tilde{\epsilon}$.

In this cost model, the duration of Hamiltonian evolution per independent sample is
\begin{equation} \label{sample_cost}
 t_{\mathrm{mix}} = \frac{1 + \widetilde{\Omega}_I}{1 - \widetilde{\Omega}_I} \sqrt{\frac{\log(1/\tilde{\epsilon})}{2}} \frac{e^{\beta \sqrt{2 (\sigma^2 + \sigma_0^2) \log(1/\tilde{\epsilon})}}}{\sigma}
\end{equation}
 with $K_E$ and truncation errors both set to $\tilde{\epsilon}$.
Because $\sigma_0$ will not be known in advance, I approximately minimize the sampling cost by minimizing it for $\sigma_0 = 0$ with the choice
\begin{equation}
 \sigma = \frac{1}{\beta \sqrt{2 \log(1/\tilde{\epsilon})}} \ \ \ \mathrm{and} \ \ \ t_{\max} = \beta \log(1/\tilde{\epsilon}).
\end{equation}
The minimal sampling cost when $\widetilde{\Omega}_I$ and $\sigma_0$ are not known in advance is then
\begin{equation}
 t_{\mathrm{mix}} = \beta \log(1/\tilde{\epsilon}) \frac{1 + \widetilde{\Omega}_I}{1 - \widetilde{\Omega}_I} e^{\sqrt{1 + 2 \beta^2\sigma_0^2 \log(1/\tilde{\epsilon})}}.
\end{equation}
Thus, the optimization of $\sigma$ can mitigate the cost of a diverging halting time for $\sigma_0 = 0$,
 but the excess variance in the halting process results in an $\Omega(\beta \sigma_0)$ asymptotic growth of $\log t_{\mathrm{mix}}$
  that persists even if $t_{\mathrm{mix}}$ is further reduced by optimizing $\sigma$ for a known $\sigma_0$.

\section{Conclusion}

In summary, the Metropolis--Hastings algorithm can be adapted to handle limited control and imprecise measurements of a state and its energy,
 which directly enables a quantum algorithm by replacing these classical SPAM operations with their quantum counterparts.
This imprecision produces a fundamental stalling problem in the delayed rejection process because a large but rare underestimate of the initial energy
 requires an equally large and rare underestimate of the final energy to accept a state and continue the Markov chain.
Stalling is a fat tail in the distribution of halting times for the delayed rejection process, and it causes the expected halting time to diverge unless rejection is halted prematurely.
Premature halting distorts the stationary state of the Markov chain away from the desired thermal state and introduces a tradeoff between cost and accuracy.

There are many ways to introduce complications into Algorithm \ref{IMH} that preserve both detailed balance and its generalization to a quantum algorithm.
For example, the simple sequence of updating observations after the first non-symmetric update
 can be replaced by a different non-symmetric update for each loop iteration that depends on all past energies and observations.
The simple acceptance probability formula in Eq.\@ (\ref{simple_accept}) will no longer be valid, but Eq.\@ (\ref{generic_accept}) can be evaluated using $O(n)$ memory at iteration $n$.
Also, each energy measurement can be repeated multiple times to construct a robust mean estimator for use in acceptance probabilities.
Finally, a biased energy estimator could be constructed using observations from previous steps of a Markov chain
 and used in rejection sampling of the initial energy measurement to avoid large energy underestimates.
Some of these features might reduce the cost of thermal state sampling in some cases, although it is not yet clear if they can reduce or repair the stalling problem in general.

I expect that a more productive research direction will be to use the lessons learned in this paper to develop a more sophisticated quantum MCMC algorithm that does not have a classical counterpart.
In particular, the delayed rejection process with SPAM operations and filtered QPE could possibly be combined with the coherent rejection process \cite{quantum_metropolis}
 that only measures one qubit to determine each state acceptance.
A SPAM operation can be implemented as a unitary operation coupled with auxiliary qubits that are only measured if a state is accepted.
If a state is rejected, then such a coherent SPAM operation can be uncomputed approximately.
A state that has a low probability of being accepted should be distorted only a small amount by the single-qubit measurement of coherent rejection,
 and thus it should be possible to uncompute the SPAM operation accurately.
There is a good chance that this mechanism can remove the fat tail from the distribution of halting times and repair the stalling problem of the quantum Metropolis--Hastings algorithm.

\begin{appendix}

\section{Evaluation of acceptance probability\label{acceptance_probability}}
\setcounter{equation}{0}\renewcommand\theequation{A\arabic{equation}}

To evaluate Eq.\@ (\ref{generic_accept}), I first define a convenient intermediate variable $B_{n,m}$ as
\begin{align} \label{B_define}
 A(\gamma_{n,m} + \beta \sigma^2) &= \min \{ 1, B_{m,n} / B_{n,m} \} , \notag \\
 B_{n \pm m,n} &= e^{-\beta \omega_n} P( o_n | o_{n \pm m} ) R(\gamma_{n \pm m,n} + \beta \sigma^2),
\end{align}
 for $\gamma_{n,m} = (o_n, \omega_n, \cdots, o_m , \omega_m) \in \mathcal{M}_{|n-m|}$.
The recursive relationship for $R$ in Eq.\@ (\ref{R_recursion}) can be rewritten as a recursive expression for $B_{n,m}$ by substituting in $B_{n,m}$ for $A$ to get
\begin{equation} \label{B_recursion}
 B_{n \pm m,n} = \left\{ \begin{array}{ll} e^{-\beta \omega_n} P( o_n | o_{n \pm 1} ), & m = 1 \\ \max \{ 0, B_{n \pm (m-1), n} - B_{n, n \pm (m-1)} \}, & m > 1 \end{array} \right. ,
\end{equation} 
 which eventually relates $B_{n,m}$ to sums over $B_{n \pm 1, n}$ at the end of the recursion.
 
To solve Eq.\@ (\ref{B_recursion}), I make use of two simplifying identities.
The first identity is
\begin{equation} \label{minmax_identity}
 \max \{ 0, \max \{ 0, \min \mathcal{X} \} + \min \{ 0, y - \min \mathcal{X} \} \} =  \max \{ 0, \min ( \mathcal{X} \cup \{ y \} ) \} 
\end{equation}
 for all $\mathcal{X} \subseteq \mathbb{R}$ and $y \in \mathbb{R}$, which is straightforward to check by considering $y \ge \min \mathcal{X}$ and $y < \min \mathcal{X}$ separately.
The second identity relates minima and maxima over sets of partial sums for two different orderings of a finite series,
\begin{align} \label{partial_sum_identity}
 \max \{ b_1, b_1 + b_2, \cdots , b_1 + \cdots + b_{n-1} \} & = b_1 + \cdots + b_n \notag \\
 & \ \ \ - \min \{ b_n, b_n + b_{n-1}, \cdots , b_n + \cdots + b_2 \},
\end{align}
 for any $n \ge 2$ and $b_1, \cdots , b_n \in \mathbb{R}$, which results from shifting every partial sum in a set by the full sum of the series and compensating the shift outside of the optimization.

Next, I evaluate the two base values of $B_{n,m}$ in the recursion,
\begin{equation} \label{B_start}
 B_{n \pm 1, n} = e^{-\beta \omega_n} P( o_n | o_{n \pm 1} ) , \ \ \ B_{n \pm 2, n} = \max \{ 0, b_n^{\pm} \} ,
\end{equation}
 which I have simplified by defining another intermediate variable,
\begin{equation} \label{b_define}
 b_n^{\pm} = e^{-\beta \omega_n} P( o_n | o_{n \pm 1} ) - e^{-\beta \omega_{n \pm 1}} P( o_{n \pm 1} | o_n),
\end{equation}
 with a redundancy, $b_n^{\pm} = - b_{n \pm 1}^{\mp}$, that helps to simplify $B_{n,m}$.

Finally, I use induction to prove a general form for $B_{n \pm m, n}$ with $m \ge 2$,
\begin{equation} \label{B_formula}
 B_{n \pm m, n} = \max \{ 0, \min \{ b_n^{\pm}, b_n^{\pm} + b_{n \pm 1}^{\pm}, \cdots , b_n^{\pm} + \cdots + b_{n \pm (m-2)}^{\pm}\} \},
\end{equation}
 beginning with $B_{n \pm 2, n}$ in Eq.\@ (\ref{B_start}).
For $m \ge 3$, I use Eq.\@ (\ref{B_recursion}) to evaluate $B_{n \pm m, n}$ from \\ $B_{n \pm (m-1), n}$ and $B_{n, n \pm (m-1)}$
 and use Eqs.\@ (\ref{partial_sum_identity}) and (\ref{b_define}) to relate their partial sums,
\begin{align}
 B_{n \pm m, n} = \max \{ 0, & \max \{ 0, \min \{ b_n^{\pm}, b_n^{\pm} + b_{n \pm 1}^{\pm}, \cdots , b_n^{\pm} + \cdots + b_{n \pm (m-3)}^{\pm}\} \} \notag \\
 & + \min \{ 0, b_n^{\pm} + \cdots + b_{n \pm (m-2)}^{\pm} \notag \\
 & \ \ \ \ \ \ \ \ \ \ \ \ \ - \min \{ b_n^{\pm}, b_n^{\pm} + b_{n \pm 1}^{\pm}, \cdots , b_n^{\pm} + \cdots + b_{n \pm (m-3)}^{\pm}\} \} \} ,
\end{align}
 which reduces to Eq.\@ (\ref{B_formula}) by Eq.\@ (\ref{minmax_identity}) and completes the inductive proof.

The form of the solution in Eq.\@ (\ref{simple_accept}) defines $B_{n \pm 1, n}$ with a modified form of Eq.\@ (\ref{B_formula})
 that includes $B_{n \pm 1, n}$ into the minimization over partial sums since $B_{n \pm 1, n} \ge b_n^{\pm}$.
Also, the numerator and denominator of $A$ are related by using Eq.\@ (\ref{partial_sum_identity}).

\section{Evaluation of halting distribution \label{halting_distribution}}
\setcounter{equation}{0}\renewcommand\theequation{B\arabic{equation}}

For the discrete halting process defined by Eq.\@ (\ref{simple_halt_process}) and $\Delta > 0$,
 I evaluate the asymptotic form of its halting distribution, tail probabilities, and expected halting time.

First, I evaluate event probabilities conditioned on $y_n$ values and marginalized over $u_n$ values.
The probability of satisfying the first halting condition is $\min \{ 1, e^{y_1-y_0-\Delta} \}$, and the probability of satisfying the $n$th halting condition for $n \ge 2$ is
\begin{equation}
 \max \left\{ 0, \min \left\{ 1, \frac{e^{y_n - y_0 - \Delta} - e^{y_n^{\max} - y_0 - \Delta}}{1 - e^{y_n^{\max} - y_0 - \Delta}} \right\} \right\}
\end{equation}
 for $y_n^{\max} = \max_{m\in \{1, \cdots , n-1\}} y_m$, on the accessible domain restricted to $y_n^{\max} < y_0 + \Delta$.
The probability of failing to satisfy all halting conditions prior to the $n$th halting condition is
\begin{equation} \label{tail0}
 (1 - e^{y_1-y_0-\Delta}) \prod_{m \in \{1, \cdots, n-1\} \setminus \{1\}}  \min \left\{ 1, \frac{1 - e^{y_m - y_0 - \Delta}}{1 - e^{y_m^{\max} - y_0 - \Delta}} \right\} = 1 - e^{y_n^{\max}-y_0-\Delta}
\end{equation}
 for $y_n^{\max} < y_0 + \Delta$, which is simplified by replacing $y_m$ with $y_{m+1}^{\max}$ and removing one from each minimized set.
For $y_n^{\max} \ge y_0 + \Delta$, this is an event with zero probability.

Next, I evaluate halting probabilities $p_{\mathrm{halt}}(n)$ from tails $t_n$ of the halting distribution,
\begin{equation} \label{halt2tail}
 p_{\mathrm{halt}}(n) = t_n - t_{n+1}, \ \ \ t_n = \sum_{m=n}^\infty  p_{\mathrm{halt}}(m).
\end{equation}
For $n \ge 2$, $t_n$ is Eq.\@ (\ref{tail0}) integrated over the distribution of $y_0$ and $y_n^{\max}$ values,
\begin{align} \label{p_tail}
 t_n = \int_{-\infty}^{\infty} d y \frac{1}{2^{n-1}} \left[ \frac{d}{dy} \mathrm{erfc} \left(  \frac{-y}{\sqrt{2\Delta}} \right)^{n-1} \right] \int_{y-\Delta}^{\infty} d y_0 \frac{e^{-y_0^2/(2 \Delta)}}{\sqrt{2 \pi \Delta}} (1 - e^{y-y_0-\Delta}),
\end{align}
 where the cumulative distribution function of $y_n^{\max}$ is the cumulative distribution function of the normal distribution of mean zero and variance $\Delta$ to the $(n-1)$th power,
 written as a complementary error function, $\mathrm{erfc}(x)$.
I simplify Eq.\@ (\ref{p_tail}) and extend it to $n \ge 1$ using integration by parts, a change of variables, and a regrouping of integrals into
\begin{equation} \label{define_s}
 t_n = s_{n} - s_{n-1}, \ \ \ s_n = \sqrt{2 \Delta} \int_{-\infty}^{\infty} dx \, e^{\sqrt{2\Delta}x - \Delta/2} \left[ 1 - \left( \frac{\mathrm{erfc}(-x)}{2} \right)^n \right],
\end{equation}
 which has explicit values $s_0 = 0$, $s_1 = 1$, and $s_2 = 2 - \mathrm{erfc} ( \sqrt{\Delta} / 2 )$.
 
I am not able to evaluate $s_n$ directly for $n \ge 3$, so I bound it instead.
I construct the bounds on $s_n$ mostly by using either $\mathrm{erfc}(x) \le e^{-x^2}$ for $x \ge 0$ or the tighter bounds
\begin{equation} \label{erfc_bound}
 0 \le \frac{e^{1 - x/y - x^2}}{\sqrt{\pi} y + 1} \le \frac{e^{-x^2}}{\sqrt{\pi} x + 1} \le \mathrm{erfc}(x) \le \frac{e^{-x^2}}{\sqrt{\pi} x} \le \frac{e^{-x^2}}{\sqrt{\pi} y} \le 2
\end{equation}
 for $x \ge y > 0$, with the outermost bounds valid for all $x \in \mathbb{R}$
 and the innermost bounds establishing the asymptotic equivalence $\mathrm{erfc}(x) \sim e^{-x^2}/(\sqrt{\pi}x)$ in the $x \rightarrow \infty$ limit.
I then use these bounds to bound $\mathrm{erfc}(-x)^n$ in the integrand of Eq.\@ (\ref{define_s}) as
\begin{equation} \label{erfc_bound2}
 \left. \begin{array}{rr} 0, & x < x_n \\ 1 - \frac{n}{2 \sqrt{\pi} x_n}e^{-x^2}, & x \ge x_n \end{array} \right\} \le \left( \frac{\mathrm{erfc}(-x)}{2} \right)^n \le \left\{ \begin{array}{ll} e^{-(2 x_n/ \sqrt{\pi}) (x - x_n)^2}, & x < x_n \\ 1, & x \ge x_n \end{array} \right.
\end{equation}
 for $x_n = \sqrt{\log(n/2)}$, with the non-trivial lower bound on $[1- \mathrm{erfc}(x)/2]^n$ from Bernoulli's inequality, $1-nz \le (1-z)^n$ for $z \le 1$, and Eq.\@ (\ref{erfc_bound}),
 and the non-trivial quadratic upper bound on $n\log(\mathrm{erfc}(-x)/2)$ from an upper bound of $-4 x_n/\sqrt{\pi}$ on its second derivative for $x \le x_n$ since it is a monotonically increasing function.
I evaluate the bounding integrals,
\begin{align} \label{s_bounds}
 [ 1 - \sqrt{\pi} \alpha_n e^{\alpha_n^2} \mathrm{erfc}(\alpha_n) ] e^{\sqrt{2\Delta} x_n - \Delta/2}  \le s_n & \le e^{\sqrt{2\Delta} x_n - \Delta/2} + \tfrac{n}{2 x_n} \sqrt{\tfrac{\Delta}{2}} \mathrm{erfc}\left(x_n - \sqrt{\tfrac{\Delta}{2}} \right)
\end{align}
 for $\alpha_n = (\sqrt[4]{\pi}/2)\sqrt{\Delta/x_n}$, and then relax and restrict them into
\begin{equation}
 ( 1 - \sqrt{\pi} \alpha_n ) e^{\sqrt{2\Delta} x_n - \Delta/2} \le s_n  \le \left( 1 +  \tfrac{1}{x_n} \sqrt{\tfrac{\Delta}{2}} \right) e^{\sqrt{2\Delta} x_n - \Delta/2}
\end{equation}
 for $x_n \ge \sqrt{\Delta/2}$.
These bounds establish an asymptotic equivalence,
\begin{equation} \label{s_equiv}
 s_n \sim e^{\sqrt{2\Delta \log n} - \Delta/2}
\end{equation}
 in the $n \rightarrow \infty$ limit, and $s_n$ also diverges in this limit.

The bounds in Eq.\@ (\ref{erfc_bound2}) can also be used to bound $t_n$ and $p_{\mathrm{halt}}(n)$ by writing them as
\begin{equation}
 r_{m,n} = \sqrt{ 2 \Delta} \int_{-\infty}^{\infty} dx \, e^{\sqrt{2\Delta}x - \Delta/2} \left(\frac{\mathrm{erfc}(x)}{2}\right)^m \left( \frac{\mathrm{erfc}(-x)}{2} \right)^n
\end{equation}
 with $t_n = r_{1,n-1}$ and $p_{\mathrm{halt}}(n) = r_{2,n-1}$.
I then apply the bounds in Eq.\@ (\ref{erfc_bound2}) to $\mathrm{erfc}(-x)^n$ in the integrand and similarly bound $\mathrm{erfc}(x)^m$ using Eq.\@ (\ref{erfc_bound}) with its base bounded as
\begin{equation}
 \left. \begin{array}{rr} 0, & x < x_n \\ \frac{1}{\sqrt{\pi} x_n + 1} e^{1 - x/x_n - x^2}, & x \ge x_n \end{array} \right\} \le \mathrm{erfc}(x) \le \left\{ \begin{array}{ll} 2 + e^{-x^2}, & x < 0 \\
 e^{-x^2}, & 0 \le x < x_n \\ \frac{1}{\sqrt{\pi} x_n} e^{-x^2}, & x \ge x_n \end{array} \right. .
\end{equation}
For $m \ge 1$ and $n \ge 3$, I integrate these relaxed bounds on $r_{m,n}$ to get
\begin{align}
 & \sqrt{\tfrac{\pi \Delta}{2}} \tfrac{e^{ m - \Delta/2}}{(2 \sqrt{\pi} x_n + 2)^m} \left( \tfrac{ e^{\beta_{m,n}^2/m} \mathrm{erfc}(\sqrt{m} x_n - \beta_{m,n}/\sqrt{m})}{\sqrt{m}} - \tfrac{ n e^{\beta_{m,n}^2/(m+1)} \mathrm{erfc}(\sqrt{m+1} x_n - \beta_{m,n}/\sqrt{m+1})}{2 x_n \sqrt{\pi(m+1)}}  \right)
 \notag \\
 &  \le r_{m,n} \le \sqrt{\tfrac{\pi \Delta}{2 m}} \tfrac{e^{(1/m - 1) \Delta/2}}{(2 \sqrt{\pi} x_n)^m} \mathrm{erfc}\left(\sqrt{m} x_n - \sqrt{\tfrac{\Delta}{2m}}\right) +
  \sqrt{\pi} \alpha_n e^{\alpha_n^2} \mathrm{erfc}(\alpha_n + \gamma_n) e^{\sqrt{2\Delta} x_n - \Delta/2} \notag \\
 & \ \ \ \ \ \ \ \ \ \ \ \ \ + \tfrac{\sqrt{\pi} \alpha_n}{ 2^m\sqrt{1 + \sqrt{\pi}/(2 x_n)}} e^{ (\alpha_n + \gamma_n)^2/[1 + \sqrt{\pi}/(2x_n)] - \gamma_n^2 -\Delta/2} \mathrm{erfc}\left( \tfrac{\alpha_n + \gamma_n}{\sqrt{1 + \sqrt{\pi}/(2 x_n)}} \right)
\end{align}
 for $\beta_{m,n} = \sqrt{\Delta/2} - m/(2 x_n)$ and $\gamma_n = \sqrt{2 x_n^3} / \sqrt[4]{\pi}$, which I further relax and restrict into
\begin{align}
 \left( 1 - \frac{1}{x_n \sqrt{\pi}} \right) \frac{ \sqrt{\Delta} e^{\sqrt{2\Delta} x_n - \Delta/2}}{(n \sqrt{\pi} x_n + n)^m (\sqrt{2} m x_n - \sqrt{2} \beta_{m,n})} \le r_{m,n} & \le \frac{ \sqrt{\Delta} e^{\sqrt{2\Delta} x_n - \Delta/2}}{(n \sqrt{\pi} x_n)^m (\sqrt{2} m x_n - \sqrt{\Delta})} \notag \\
 & \ \ \ + \frac{2 \alpha_n e^{-\gamma_n^2 - \Delta/2}}{\alpha_n + \gamma_n}
\end{align}
 for $m x_n > \sqrt{\Delta/2}$.
These bounds establish another asymptotic equivalence,
\begin{equation}
 r_{m,n} \sim \sqrt{\frac{\Delta}{2 \log n}} \frac{e^{\sqrt{2\Delta\log n} - \Delta/2}}{m (n \sqrt{\pi \log n})^m}
\end{equation}
 in the $n \rightarrow \infty$ limit, or simply $r_{m,n} = \omega(n^{-m})$ in Bachmann--Landau notation.

The expected halting time when the process is limited to a maximum of $n$ steps is
\begin{equation} \label{halt2s}
 \tilde{n}_{\mathrm{halt}}(n) = \sum_{m=1}^{n-1} m \, p_{\mathrm{halt}}(m) +  n \, t_n = s_n,
\end{equation}
and the truncation error $\tilde{\epsilon}$ in Eq.\@ (\ref{define_error}) for an idle update corresponds to $t_{n+1}$.

\end{appendix}

\newpage

\begin{acknowledgements}
J. E. M. thanks Andrew Baczewski for useful discussions.
The Molecular Sciences Software Institute is supported by grant CHE-2136142 from the National Science Foundation.
\end{acknowledgements}


\begin{thebibliography}{99}
\bibitem{quantum_simulation} Richard P. Feynman. ``Simulating physics with computers''. \href{https://doi.org/10.1007/BF02650179}{Int. J. Theor. Phys. \textbf{21}, 467--488} (1982).
\bibitem{mcmc} David P. Landau and Kurt Binder. ``A Guide to Monte Carlo Simulations in Statistical Physics''. Cambridge University Press, Cambridge (2015).
\bibitem{metropolis} Nicholas Metropolis, Arianna W. Rosenbluth, Marshall N. Rosenbluth, Augusta H. Teller, and Edward Teller. ``Equation of State Calculations by Fast Computing Machines''. \href{https://doi.org/10.1063/1.1699114}{J. Chem. Phys. \textbf{21}, 1087--1092} (1953).
\bibitem{metropolis_history} J. E. Gubernatis. ``Marshall Rosenbluth and the Metropolis algorithm''. \href{https://doi.org/10.1063/1.1887186}{Phys. Plasmas \textbf{12}, 057303} (2005).
\bibitem{hastings} W. K. Hastings. ``Monte Carlo sampling methods using Markov chains and their applications''. \href{https://doi.org/10.1093/biomet/57.1.97}{Biometrika \textbf{57}, 97--109} (1970).
\bibitem{hastings_history} Christian Robert and George Casella. ``A Short History of Markov Chain Monte Carlo: Subjective Recollections from Incomplete Data''. \href{https://doi.org/10.1214/10-STS351}{Statist. Sci. \textbf{26}, 102--115} (2011).
\bibitem{qmcmc0} Barbara M. Terhal and David P. DiVincenzo. ``Problem of equilibration and the computation of correlation functions on a quantum computer''. \href{https://doi.org/10.1103/PhysRevA.61.022301}{Phys. Rev. A \textbf{61}, 022301} (2000).
\bibitem{qmcmc1} Jonathan E. Moussa. ``Low-Depth Quantum Metropolis Algorithm'' (2022). \href{https://arxiv.org/abs/1903.01451}{arXiv:1903.01451}.
\bibitem{qmcmc2} Chi-Fang Chen, Michael J. Kastoryano, Andr\'{a}s Gily\'{e}n. ``An efficient and exact noncommutative quantum Gibbs sampler'' (2023). \href{https://doi.org/10.48550/arXiv.2311.09207}{arXiv:2311.09207}.
\bibitem{qmcmc3} Zhiyan Ding, Bowen Li, and Lin Lin. ``Efficient Quantum Gibbs Samplers with Kubo--Martin--Schwinger Detailed Balance Condition''. \href{https://doi.org/10.1007/s00220-025-05235-3}{Commun. Math. Phys. \textbf{406}, 67} (2025).
\bibitem{qmcmc4} Andr\'{a}s Gily\'{e}n, Chi-Fang Chen, Joao F. Doriguello, and Michael J. Kastoryano. ``Quantum generalizations of Glauber and Metropolis dynamics'' (2024). \href{https://doi.org/10.48550/arXiv.2405.20322}{arXiv:2405.20322}.
\bibitem{quantum_metropolis} K. Temme, T. J. Osborne, K. G. Vollbrecht, D. Poulin, and F. Verstraete. ``Quantum Metropolis sampling''. \href{https://doi.org/10.1038/nature09770}{Nature \textbf{471}, 87--90} (2011).
\bibitem{delayed_rejection} Antonietta Mira. ``On Metropolis-Hastings algorithms with delayed rejection''. \href{https://econpapers.repec.org/RePEc:mtn:ancoec:2001:3:16}{Metron \textbf{59}, 231--241} (2001).
\bibitem{swendsen_wang} Robert H. Swendsen and Jian-Sheng Wang. ``Nonuniversal critical dynamics in Monte Carlo simulations''. \href{https://doi.org/10.1103/PhysRevLett.58.86}{Phys. Rev. Lett. \textbf{58}, 86--88} (1987).
\bibitem{uncertain_metropolis} D. M. Ceperley and M. Dewing. ``The penalty method for random walks with uncertain energies''. \href{https://doi.org/10.1063/1.478034}{J. Chem. Phys. \textbf{110}, 9812--9820} (1999).
\bibitem{qpe} Daniel S. Abrams and Seth Lloyd, ``Quantum Algorithm Providing Exponential Speed Increase for Finding Eigenvalues and Eigenvectors''. \href{https://doi.org/10.1103/PhysRevLett.83.5162}{Phys. Rev. Lett. \textbf{83}, 5162--5165} (1999).
\bibitem{quantum_pseudocode} E. Knill, ``Conventions for Quantum Pseudocode'' (2022). \href{https://doi.org/10.48550/arXiv.2211.02559}{arXiv:2211.02559}.
\end{thebibliography}
\end{document}